\documentclass[]{jfm}

\usepackage{graphicx}
\usepackage[justification=justified,format=plain, width=\textwidth]{caption}
\usepackage{newtxtext}
\usepackage{newtxmath}
\usepackage{natbib}

\usepackage{hyperref}
\hypersetup{
    colorlinks = true,
    urlcolor   = blue,
    citecolor  = blue,
}

\newcommand{\RomanNumeralCaps}[1]
\linenumbers
\usepackage{array}
\usepackage{epstopdf}
\newcolumntype{L}{>{\centering\arraybackslash} m{3cm}}

\title{Multilayer network analysis to study complex inter-subsystem interactions in a turbulent thermoacoustic system}

\author{Shruti Tandon \aff{1} and R. I. Sujith \aff{1}
  \corresp{\email{sujith@iitm.ac.in}}}

\affiliation{\aff{1}Department of Aerospace Engineering, Indian Institute of Technology Madras, Chennai 600 036, India}

\begin{document}
\maketitle

\begin{abstract}
Thermoacoustic systems are complex systems where the interactions between the hydrodynamic, acoustic and heat release rate fluctuations lead to diverse dynamics such as chaos, intermittency, and ordered dynamics. Such complex interactions cause catastrophically high-amplitude acoustic pressure oscillations and the emergence of order in the spatio-temporal dynamics, referred to as thermoacoustic instability. In this work, we use multilayer networks to study the spatial pattern of inter-subsystem interactions between the vorticity dynamics and thermoacoustic power generated due to acoustically-coupled combustion in a bluff-body stabilized turbulent dump combustor. We construct a two-layered network where the layers represent the thermoacoustic power and vorticity fields. The inter-layer links are determined using cross-variable short-window correlations between vorticity and thermoacoustic power fluctuations at any two locations in the flow field. Analyzing the topology of inter-layer networks, using network properties such as degree correlations and link-rank distributions, helps us infer the spatial inhomogeneities in inter-subsystem interactions and unravel the fluid mechanical processes involved during different dynamical states. We show that, during chaotic dynamics, interactions between subsystems are non-localized and spread throughout the flow field of the combustor. During the state of thermoacoustic instability (order), we find that intense interactions occur in between regions of coherent vortex shedding and thermoacoustic power generation and we understand that these processes are strongly and locally coupled. Moreover, we discover that such dense inter-layer connections emerge in spatial pockets in the dump plane of the combustor during the state of intermittency much prior to the onset of order.
\end{abstract}

\section{Introduction}

Thermoacoustic systems, such as those in combustion chambers of gas-turbine or rocket engines, can exhibit diverse dynamical states owing to nonlinear interactions between the turbulent hydrodynamic field, the acoustic pressure oscillations and the heat released due to combustion \citep{lieuwen2005combustion, sujith2020complex, sujith2021thermoacoustic}. A thermoacoustic system is a viscous thermo-fluid system where combustion occurs in a closed duct coupled with an acoustic field that is established in the duct. Such a system comprises three modes of propagation acoustic, vorticity and entropy modes \citep{chu1958non}. (Entropy modes represent the production and convection of hot spots of combustion in viscous flows.) In a homogeneous uniform flow, these modes are decoupled and propagate independent of each other. However, if the underlying flow is turbulent, these modes are nonlinearly coupled leading to self-sustained feedback between these modes. For example, the interaction between vorticity and acoustic modes leads to the production of both sound and vorticity \citep{chu1958non, lieuwen2021unsteady}.

Thermoacoustic systems are essentially complex systems that exhibit self-organized dynamics \citep{gershenson2003can}. A system is considered to be \textit{complex} if it comprises multiple subsystems and the dynamics of the system is determined by the interaction between these constituent subsystems \citep{ottino2003complex, holovatch2017complex}. Dynamics of complex systems cannot be fully comprehended by studying any of the subsystems in isolation from the other subsystems as is done in conventional reductionist approaches. Many problems in fluid mechanics pose similar challenges. For example, turbulence which has vortical interactions at multiple timescales, multi-phase flows containing interactions between the different phases (such as fluid and particles), and the climate system where interactions between wind flow, cloud cover, solar radiation and the earth's topology together determine the dynamics \citep{rind1999complexity}. In such systems, it is essential to analyse the spatio-temporal interactions both within each subsystem and between subsystems.

\par The pattern of intra and inter-subsystem interactions and the resulting dynamics are sensitive to the control parameters of the system. In a thermoacoustic system, the pattern of nonlinear interactions between the acoustic, hydrodynamic and combustion subsystems varies with the inlet mass flow rates of fuel and air (or Reynolds number), as well as the ratio of these fuel and air mass flow rates (or equivalence ratio). The stable state of operation of a combustor is often referred to as combustion noise. This state is characterized by chaotic spatio-temporal dynamics such as incoherent heat release and irregular shedding of small vortices \citep{george2018pattern} and low-amplitude aperiodic fluctuations in the acoustic pressure signal \citep{gotoda2011dynamic, nair2014intermittency}. Moreover, the acoustic pressure time series obtained during the state of combustion noise delineates high-dimensional chaos \citep{tony2015detecting}. With change in the control parameters, the complex feedback interactions between the subsystems of a thermoacoustic system cause more acoustic driving than acoustic damping in the system. As a result, the system exhibits thermoacoustic instability which is characterized by ordered dynamics including large-amplitude periodic oscillations in the acoustic pressure signal \citep{lieuwen2002experimental}, coherent heat release and periodic shedding of large coherent structures (vortices) \citep{george2018pattern, poinsot1987vortex}. The transition from combustion noise (chaos) to thermoacoustic instability (order) occurs via the route of intermittency which is characterized by epochs of periodicity interspersed by epochs of chaotic dynamics \citep{nair2014intermittency, unni2017flame, kheirkhah2017dynamics, ebi2018flame}.

\par The occurrence of thermoacoustic instability is catastrophic and can lead to humongous loss of resources and revenue. Several studies have been invested on understanding the phenomenological causes and devising effective means of control of thermoacoustic instability \citep{candel2002combustion, lieuwen2005combustion, sujith2021thermoacoustic}. For example, researchers use different approaches such as dynamical systems and bifurcation theory \citep{ananthkrishnan1998application, lieuwen2002experimental, gotoda2014detection, laera2017flame}, framework of synchronization \citep{pawar2017thermoacoustic, mondal2017onset}, measuring flame response to the acoustic field perturbations using impulse response functions \citep{polifke2020modeling} or flame transfer/ describing functions \citep{dowling1997nonlinear, noiray2008unified, hemchandra2011heat, schuller2020dynamics}, and complex networks \citep{murugesan2015combustion, gotoda2017characterization, sujith2020complex}. Recently, the transition from combustion noise to thermoacoustic instability has been viewed as a transition from chaos to order in the turbulent reacting flow field of a thermoacoustic system \citep{tandon2021condensation, sujith2021thermoacoustic}. However, to fully comprehend the emergence of order in the spatio-temporal dynamics of a thermoacoustic system, we must investigate the pattern of nonlinear interactions between multiple constituent subsystems during various dynamical states. For example, we must study how the emergence of periodicity in the acoustic field may be related to the emergence of large vortices in the combustor; or how the spatio-temporal distribution of heat release rate fluctuations are dependent on the acoustic and the hydrodynamic fields.

\par Early seminal works by \citet{poinsot1987vortex}, \citet{sterling1987longitudinal} and \citet{schadow1992combustion} discuss the role of coherent structures in exciting periodic dynamics in the combustion chamber, thus leading to thermoacoustic instability. In these early attempts, techniques such as spark-Schlieren imaging and C$_2$ radiation maps were used to investigate the simultaneous evolution of large coherent structures and the motion of regions of intense heat release, respectively. These studies describe the complete cycle of formation, growth and decay of large coherent structures (vortices) and the burning of fuel-air mixture carried by these vortices. The authors show that combustion occurs in the wake of vortices and the reaction zones trail behind the vortices. 

\par Vortices cause mixing of the jet of fresh mixture and the burnt hot products in the flow field. Further, these large coherent structures promote bulk mixing, while fine-scale mixing of hot products and unburnt reactants occurs along the braids of such vortices (regions of high strain rate), or in regions of high velocity gradients and where tendency of rotation of local fluid particles is high \citep{schadow1992combustion}. Also, vortices may impinge onto the walls of the combustor and break down causing fine-scale mixing of fuel and air. When sufficient mixing occurs, all the fuel-air mixture contained in the large vortex undergoes combustion at once, thus causing large and sudden heat release in the combustion chamber. A feedback between the acoustic velocity fluctuations and the vorticity field leads to self-sustained periodic shedding of  large vortices subsequently causing periodic spikes in the heat release rate fluctuations. In locations where the heat release rate fluctuations are in phase with the acoustic pressure, acoustic driving becomes dominant eventually leading to thermoacoustic instability \citep{rayleigh1878explanation, sterling1987longitudinal}.

\par Thus, different studies have theorized that there are inter-dependencies between the acoustic modes, coherent structures and heat release rate fluctuations during the state of thermoacoustic instability. Subsequently, the role of acoustic modes on the evolution of coherent structures during the state of thermoacoustic instability has been investigated using different techniques such as phase averaging and proper orthogonal decomposition \citep{lacarelle2010combination, oberleithner2011three, tammisola2016coherent}, and dynamic mode decomposition \citep{schmid2011applications, premchand2019lagrangian}, and also using  numerical models \citep{matveev2003model}. Moreover, active control techniques have been developed by modifying vortex shedding frequencies and suppressing the coherence of large vortices that aid in exciting thermoacoustic instability \citep{mcmanus1993review, paschereit1999coherent}. However, the understanding about inter-dependencies between acoustic, vorticity and combustion dynamics during the states of combustion noise or intermittency is not known and that during the state of thermoacoustic instability is far from complete.

Due to inter-subsystem interactions between the acoustic, hydrodynamic and combustion dynamics, a self-sustained feedback loop is generated wherein the evolution of each subsystem is dependent on the other. Note that, a thermoacoustic system is a spatially-extended system where nonlinear interactions compete with turbulence to influence the dynamics. Moreover, there are time delays associated with vortex formation and convection, and the mixing and combustion of the fuel-air mixture contained in these vortices. As a result, the strength of inter-subsystem interactions will vary in space and time inside the combustor. Strong interactions can occur in local pockets or across spatially separated regions throughout the flow field. Also, time-delayed interactions can occur in between physically separated regions due to the convection of vortices. Thus, inter-subsystem interactions are spatially inhomogeneous and it is essential to characterize such inhomogeneities in the spatial pattern of interactions to interpret the underlying physics. An interesting question is, how does the spatial pattern of inter-subsystem interactions differ during various dynamical states such as chaos (combustion noise), intermittency and order (thermoacoustic instability)? Answering such questions can aid in understanding the emergence of order amidst chaos in the spatio-temporal dynamics of a thermoacoustic system, and distinguish the fluid mechanical processes during distinct dynamical states.

\par Complex systems, such as thermoacoustic systems, are best analysed using complex networks. A complex network is a set of nodes connected by links, where these links are defined by the relation between the nodes. Various network construction techniques can be used to construct networks from temporal or spatio-temporal data to infer the dynamics of that system \citep{gao2017complex, iacobello2021review}. Networks have been used extensively to characterize climate systems \citep{tsonis2006networks}, fluid mechanics problems involving turbulence \citep{iacobello2018spatial,taira2016network}, and thermoacoustic systems \citep{sujith2020complex}. 

\par In thermoacoustic literature, different types of complex networks have been used to investigate the transition between diverse dynamical states in a combustor. For example, complex networks derived from the time series of acoustic pressure have been used to analyse the emergence of periodicity in the acoustic field of a thermoacoustic system. Complex networks based on visibility algorithm \citep{murugesan2015combustion} revealed that the scale-free nature of networks derived from the acoustic pressure dynamics of a thermoacoustic system is lost as order emerges amidst chaos in the dynamics of the system. Further, recurrence analysis of acoustic pressure oscillations captures transitions of the dynamical states of a turbulent combustor and offers early warning signals for thermoacoustic instability \citep{godavarthi2017recurrence}. Also, joint recurrence networks were used to analyse the interdependence between the time series of acoustic pressure and heat release rate fluctuations during different dynamical states in a turbulent combustor \citep{godavarthi2018coupled}. Variation of measures derived from visibility network and recurrence networks were used to detect the onset of thermoacoustic instability as well as flame-blowout \citep{gotoda2017characterization}. 

\par Another network construction method based on correlation between phase space cycles derived from acoustic pressure time series highlighted the psuedo-periodicity and high-dimensional nature of acoustic pressure dynamics during the state of thermoacoustic instability \citep{okuno2015dynamics}. Further, \citet{tandon2021condensation} used cycle networks to show that as order emerges from chaos in the acoustic pressure dynamics of a turbulent combustor, the phase space topology morphs from a set of several highly unstable periodic orbits during the occurrence of chaotic dynamics (combustion noise), to a combination of moderately stable and unstable periodic orbits during intermittency, and finally to a stable limit cycle attractor during ordered dynamics (thermoacoustic instability). Also, \citet{kobayashi2019early} have used ordinal partition transition networks based on the synchronization of acoustic pressure and heat release rate fluctuations to identify the onset of thermoacoustic instability. Moreover, \citet{aoki2020dynamic} used the network entropy of ordinal partition transition networks built from the acoustic pressure signal to identify the transition between chaotic and periodic dynamics during the occurrence of intermittency.

\par Further, transitions in spatio-temporal dynamics of thermoacoustic systems have also been investigated using complex networks. Networks are constructed by treating spatial points in the flow field as nodes. The links between these nodes are determined via some relation between the variation of a certain flow variable, such as velocity or heat release rate fluctuations, at the location of these nodes. Here, the relation between nodes may be defined via linear or nonlinear measures such as correlation, mutual information or Biot-Savart law for induced velocity \citep{taira2016network}. For instance, \citet{unni2018emergence} used correlation between velocity fluctuations at different locations to construct unweighted networks. Using these time-averaged spatial networks, the authors identify the different regions where such correlations are most significant during distinct dynamical states. Similarly, \citet{krishnan2019mitigation} constructed weighted networks from velocity correlations and suggested passive control strategies to mitigate thermoacoustic instability by targeting regions of high node strengths in the velocity-based networks. Recently, \citet{hashimoto2019spatiotemporal} built turbulence networks where the links are determined by the velocity induced at a certain location due to vorticity at another location using the Biot-Savart law \citep{taira2016network}. Eventually, \citet{kurosaka2021attenuation} and \citet{abin2021jfmtaira} showed that attacking the spatial locations of primary hubs identified by the turbulence network using localized secondary jet injections can mitigate the occurrence of thermoacoustic instability. Also, complex networks were constructed from the spatial field of thermoacoustic power generated in a turbulent combustor to analyse the feedback between the heat released due to combustion and the acoustic field \citep{abin2019jfm, shima2021formation}.

\par Note that, in the literature hitherto, networks are constructed using the information of a single dynamic variable derived from the turbulent reacting flow field of a combustor. Thus, such a network construction represents the interactions within one subsystem of the combustor on what is called a `single-layer' network. For example, a network built from the velocity fluctuations represents the spatio-temporal interactions within the hydrodynamic field alone. Such single-subsystem (or single-layer) networks do not encode inter-subsystem interactions. However, thermoacoustic systems have multiple levels of complexity, that is, complex interactions occur across spatial locations within subsystems as well as between subsystems. The diverse dynamics observed in a thermoacoustic system are clearly emergent phenomena arising due to the co-evolution of and interactions between the hydrodynamic, acoustic and combustion fields \citep{sujith2020complex}. Further, the evolution of each subsystem also depends on the evolution of the other subsystems. It is essential to investigate the mutual dependencies between subsystems and between dynamics at different locations in order to comprehend the emergence of spatio-temporal patterns in the combustor. One natural solution to further decipher the higher-level complexity of such systems is to investigate the dynamics through a network of single-layer networks, that is, a multilayer network.

\par A multilayer network consists of layers representing different subsystems of a complex system \citep{bianconi2018multilayer}. Links established between nodes within a layer represent interactions within that subsystem, while links between nodes in different layers represent the interactions between different subsystems. Note that, in a complex system, the term `interaction' represents mutual inter-dependency or co-evolution of dynamics between two constituent entities of the system \citep{bertalanffy1968general, kauffman1995home, heylighen2006complexity}. The structure of inter-connections and the underlying physical processes are inter-dependent \citep{arshinov2003causality, fuchs2003structuration}. One cannot assume that the web of interactions is a result of the physical processes; rather the pattern of interactions alters the physical process as much as the underlying physics determines the structure of interactions across multiple subsystems and locations \citep{fuchs2003structuration}. In a complex system with large number of interactions occurring across multiple constituent entities (or locations), it is equivalent to study either the set of physical processes that ensue simultaneously, the visible pattern in the dynamics or the structure of the web of interactions \citep{hardy2001self, witherington2011taking}. Thus, we use multilayer networks to construct the web of interactions across subsystems and locations in a turbulent thermoacoustic system. We then study the structure of this web of interactions to infer the underlying physical processes.

Multilayer networks have been discussed extensively and well-developed mathematically in recent years \citep{boccaletti2014structure, kivela2014multilayer,domenico2013mathematical,bianconi2018multilayer, aleta2019multilayer}, with some practical applications in complex biological networks \citep{de2017multilayer}, ecological networks \citep{finn2019use}, social networks \citep{barrett2012taking,murase2014multilayer,kang2022measuring} and climate networks \citep{ying2020rossby,donges2011investigating}. Multilayer networks are discussed in several forms and by various names in the literature. Generally, multilayer networks are classified as multiplex \citep{domenico2013mathematical, boccaletti2014structure} and multi-slice networks \citep{bianconi2018multilayer}, network of networks or interconnected networks, etc. \citep{boccaletti2014structure, bianconi2018multilayer}. 

\par The application of multilayer networks to fluid mechanics can offer great insight and help reveal deeper complexities in fluid mechanics problems such as those related with multiple phases, interaction of numerous vortical structures, and multiple timescales. Specific to thermoacoustic systems, the approach involving multilayer networks is the key to answering the questions we have raised previously. We construct a two-layered network where one layer represents the hydrodynamic subsystem and the other layer represents the dynamics of the acoustically-coupled heat release rate field in the combustor. In order to decipher the structure of inter-subsystem interactions during various dynamical states, we analyse how the links are distributed between different spatial locations across the two layers of the combustor. To do so, we introduce the concept of inter-layer network assortativity and inter-layer link-rank distribution.

\par Our analysis using multilayer networks, unravels the interactions between the acoustic, combustion and hydrodynamic subsystems in a turbulent thermoacoustic system. We are able to derive insight into and distinguish the physical processes involved during distinct dynamical states in the system and our results corroborate with conjectures in earlier works \citep{chu1958non, poinsot1987vortex, schadow1992combustion}. Moreover, we describe the spatial inhomogeneties associated with inter-subsystem interactions in the flow field of the combustor during each dynamical state, which are not known in the literature hitherto. We show that spatial pockets of intense inter-subsystem interactions emerge during the occurrence of intermittency as well as thermoacoustic instability in the combustor.

\par The details of the experimental setup and data processing are discussed in Section \ref{sec_expt}, while the steps involved in the inter-layer network construction are presented in Section \ref{sec_ntwkconstruct}. Network properties such as assortativity and link-rank distribution for multilayer networks are discussed in Section \ref{sec_tools_multilayer}. The results from multilayer network analysis of different dynamical states in a turbulent bluff-body stabilized dump combustor are discussed in Section \ref{sec_results}. We conclude the study and discuss the scope for future work in Section \ref{sec_conclusion}.
\section{Experiments in a turbulent thermoacoustic system\label{sec_expt}}
\par
\begin{figure}
    \centering
    \includegraphics[width=1\linewidth]{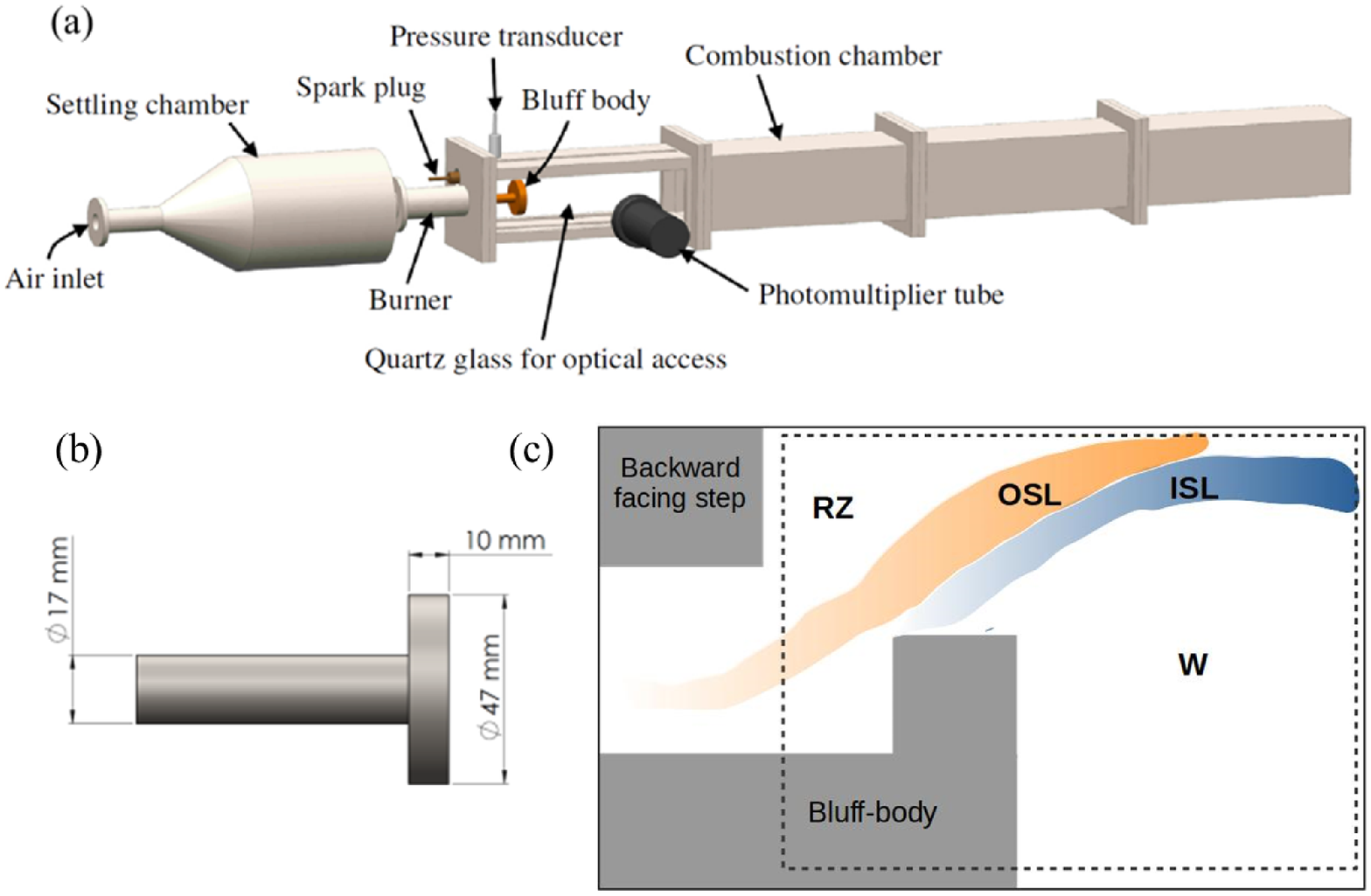}
    \caption{Schematic of (a) the experiment set-up of a turbulent combustor and (b) circular bluff-body used for flame stabilization. (c) Diagram (not to scale) of the turbulent combustor showing different regimes, namely the recirculation zone (RZ) behind a backward-facing step, also called dump plane, the outer shear layer (OSL), the inner shear layer (ISL) and the wake (W) of the bluff-body. The dotted line shows the analysis window considered in the current work. The flow is from left to right. When the flow turns around the backward-facing step, anti-clockwise vortices are shed in the recirculation zone and the outer shear layer. Due to the flow turning around the bluff-body, clockwise vortices are shed from the tip of the bluff body constituting the inner shear layer. (Figures (a) and (b) are reproduced with the kind permission from \citet{pawarphd}, Ph. D. thesis, copyright Indian Institute of Technology Madras.)}
    \label{fig_schematic}
\end{figure}

 Experiments were performed in a turbulent combustor with a backward-facing step and a bluff-body to stabilize the flame for combustion. A schematic of the experimental setup is shown in figure \ref{fig_schematic}. The cross-section of the combustion chamber is $90~\text{mm}\times 90~\text{mm}$ while its length is $1100~ \text{mm}$. The bluff-body used as a flame holding device is a circular disk of $10$ mm thickness and $47$ mm diameter located at a distance of $4.5$ mm from the inlet. The bluff-body is mounted on a central shaft having a diameter of $16$ mm. The fuel is provided through a $16$ mm diameter central shaft and ejected through four circumferential holes of $1$ mm diameter. Fuel (liquid petroleum gas) and air are mixed in the burner before being inlet into the combustion chamber. This fuel-air mixture is injected at a distance of $110$ mm upstream of the dump plane and ignited using a spark plug. The variation of fuel and air flow rates is measured in standard litres per minute (SLPM) and is monitored using mass flow controllers (Alicat Scientific, MCR Series). In the experiments, the mass flow rate of fuel is maintained at $\dot{m_f}=30 \pm 0.44$ SLPM while the mass flow rate of air ($\dot{m_a}$) is varied quasi-statically between $480\pm 7.84$ to $780\pm 10.24$ SLPM. As a result, the Reynolds number ($Re$) and the global equivalence ratio ($\phi$) of the inlet fuel-air mixture also vary with subsequent uncertainties of $\pm 6\%$ and $\pm 0.02$, respectively. The Reynolds number and the equivalence ratio ($\phi$) are calculated based on the procedure given in \citet{nair2014intermittency}, taking into account the mixture viscosity \citep{wilke1950viscosity}. The range of Reynolds numbers and equivalence ratio reported in the experiment are $18000-24000$ and $0.99-0.60$, respectively. 
\par The temporal variation of acoustic pressure is measured using a piezoelectric pressure transducer (PCB 103B02 having a sensitivity of $223.4$ mV/kPa and measurement uncertainty is $\pm 0.15$ Pa) The time series of global heat release rate is acquired using a photomultiplier tube (Hamamatsu H10722-01) fitted with an  $\text{OH}^*$ filter (narrow bandwidth filter centred at $308$ nm and $12$ nm full width at half maximum). Both the time series are acquired at a sampling rate of $10$ kHz for $3$ s. Simultaneous high-speed particle image velocimetry (PIV) and $\text{OH}^*$ chemiluminescence imaging are performed to acquire velocity field measurements and the spatio-temporal variation of unsteady heat release rate, respectively. A high-speed CMOS camera (Phantom-v 12.1) outfitted with an $\text{OH}^*$ filter (308 nm, 12 nm FWHM) is used to acquire the $\text{OH}^*$ chemiluminescence images of the flame at a frequency of 2000 frames per second for 1.258 s for each equivalence ratio. The physical region captured has dimensions $67~\text{mm} \times 44 ~\text{mm}$  which is mapped to $800 \times 600$ pixels resolution. 

\par For PIV, the region of interest is illuminated by a double pulsed Nd:YLF laser ($527$ nm, $25$ mJ at 1 kHz frequency, Photonics laser). A high-speed CMOS camera (Photron FASTCAM SA4) equipped with a ZEISS $100$ mm lens along with a short bandpass optical filter centred at $527$ nm ($12$ nm FWHM) is used to capture the light scattered (Mie scattering) by the seeding particles ($1~\mu m$ TiO$_2$ particles). We use a fluidized bed seeder for an adequate supply and uniform distribution of TiO2 in the flow. The detailed procedure for optimizing the seeding control is described in \citet{raghunathan2020multifractal}. The initial resolution of the Mie scattering images acquired using the Photron camera is $1024 \times 1024$ pixels at a frequency of 2000 Hz covering a physical region of dimensions $58.5~\text{mm} \times 58.5 ~\text{mm}$. To obtain the velocity and vorticity field data, a cross-correlation algorithm with multiple pass grid refining technique is used on the Mie scattering images. The size of the interrogation window and the maximum overlap is chosen such that, for all flow rates, the resolution of the velocity field remains the same. After numerous post-processing steps as discussed by \citet{george2018pattern}, the size of the region of interest becomes $41 \times 41$ and $63 \times 63$ spatial points for the data acquired at inlet air mass flow rates greater than and lesser than $600$ SLPM, respectively. The velocity field data acquired after post-processing of the PIV data is obtained at a frequency of $1000$ Hz for $1.36$ s. For further details of the analysis tools used to extract and process the experimental data and the uncertainty of velocity measurement, the reader may refer to \citet{george2018pattern}. 

\par In our experiments, at low inlet mass flow rates and high equivalence ratio, the combustor operates in stable mode. This state is called combustion noise and is characterized by low-amplitude aperiodic acoustic field fluctuations. As we increase the inlet mass flow rate of air (and subsequently decrease the global equivalence ratio), the combustor exhibits the state of intermittency characterized by short epochs of high-amplitude periodic oscillations amidst low-amplitude aperiodic fluctuations in the acoustic field. With further increase in the inlet mass flow rate of air (decrease in the global equivalence ratio), the combustor operates in unstable mode, referred to as thermoacoustic instability, which is characterized by high-amplitude periodic oscillations in the acoustic field.  
\section{Constructing the inter-layer network from experimental data \label{sec_ntwkconstruct}}

\par To construct multilayer networks, we consider two different physical quantities derived from the experimental data, namely vorticity ($\omega$) and $p'\dot{q}'$ fluctuations. A layer of the network represents a subsystem and each layer consists of the same set of nodes representing the same spatial points in the flow field. Here, $p'\dot{q}'$ is the temporal product of acoustic pressure ($p'$) and the heat release rate fluctuations ($\dot{q}'$) at any specific location. Note that $p'\dot{q}'$ is an important physical quantity which on being integrated over a time period and region represents the thermoacoustic power generated in that region. As explained by Lord \citet{rayleigh1878explanation}, heat released due to combustion in a region adds energy to the acoustic field when the time average of $p'\dot{q}'$ is greater than zero in that region, and the dynamics in this region is said to cause acoustic driving (or acoustic damping if $p'\dot{q}'<0$). Thus, examining the $p'\dot{q}'$-field helps us identify spatial inhomogeneties in driving/ damping mechanisms arising due to the heat released during combustion. 

\par Furthermore, combustion occurs when reactants are sufficiently mixed with hot products at the molecular level. Such fine-scale mixing of reactants and products occurs in regions with strong velocity gradients and due to local rotation of fluid parcels \citep{schadow1992combustion}. In order to identify such regions that facilitate fine-scale mixing and combustion, we use vorticity field data. Thus, we construct a two-layered network using $p'\dot{q}'$-field (comprising the thermoacoustic power layer or $p'\dot{q}'$-layer) and $\omega$-field (comprising the vorticity layer or $\omega$-layer) to decipher the inter-dependence between the vortex structure and dynamics and the thermoacoustic power generated due to acoustically-coupled combustion in a thermoacoustic system. Next, an inter-layer link is established between a node in the $p'\dot{q}'$-layer and a node in the $\omega$-layer if the time series of $p'\dot{q} '$ and $\omega$ at the respective locations are correlated over short epochs (as explained later). The resulting network is hereafter referred to as inter-layer network.

\par In order to construct such a network, we make the spatial and temporal resolution of the $\omega$ and $p'\dot{q}'$ fields equal. We compare the analysis windows of the cameras used for chemiluminescence imaging and PIV, and then select the common region with dimensions $58.5~\text{mm} \times 58.5 ~\text{mm}$ with a common spatial resolution of $39 \times 39$ pixels. Data derived from chemiluminescence imaging during all states, and the processed PIV data for inlet air flow rates less than $600$ SLPM have higher resolution than that for air flow rates more than $600$ SLPM, which is reduced to $39 \times 39$ pixels using bicubic interpolation method \citep{zhang2011interpolation}. As a result, we have the same number of nodes at precisely the same spatial locations in both the layers. In multilayer network terminology, nodes representing the same entity (here, spatial location) in different layers are referred to as replica nodes \citep{bianconi2018multilayer}. We extract the acoustic pressure, vorticity and heat release rate fluctuations at a common temporal resolution of $1000$ Hz frequency for $1.25~s$. After excluding the spatial points that lie in the region of the bluff-body location, we obtain $N= 1037$ nodes containing non-trivial data in each layer. 

\par To calculate the correlation between two time series we use an adjusted correlation function $\mathscr{R}_a$, which is a variant of the conventional Pearson correlation measure. Note that, if two time series do not have similar probability distributions, the range of their correlation may be distorted from the theoretically expected range of $[-1,1]$ and may be smaller than this ideal range \citep{ratner2009correlation}. To correct for the range, we find the maximum and minimum possible correlation between the two time series. Here, the data points of both the input time series are rearranged in ascending order of their values and a correlation is found, called the positive rematch which is essentially the maximum possible correlation. Similarly, if the data of one of the input time series is rearranged in ascending and that of the other in descending order while finding correlation, we obtain the minimum possible correlation called the negative rematch. The actual range of correlation of the two input time series is between the negative and the positive rematch. 

\par Note that, we need to compare the correlation values obtained from correlating the time series of different physical variables for various pairs of locations. To make a fair comparison these correlation values must lie in the same range (that is [-1,1]) for each pair of locations. In order to correct the range of correlation and bring it to the ideal range, we normalise the Pearson correlation value, as suggested by \citet{ratner2009correlation}. If the conventional Pearson correlation between the two time series is positive (or negative), we normalise it with the positive (or negative) rematch; and the correlation value thus obtained is called the `adjusted' correlation. Further, note that we consider the absolute value of correlations, since both highly positive and negative correlations represent co-evolution and interaction between dynamics at the two locations. The advantage of using normalised correlation values is described in detail in Appendix \ref{App_adjcorr}.

\par Further, the timescale at which the dynamics ensues in the hydrodynamic field is not necessarily the same as that in the acoustic and the combustion subsystems, especially during the states of combustion noise and intermittency. Due to distinct timescales involved, the calculation of correlation between $p'\dot{q} '$ and $\omega$ fluctuations is not straightforward. The correlations within the hydrodynamic field may be sustained only for short epochs (a few cycles of the dominant acoustic mode) owing to turbulent fluctuations \citep{lieuwen2002experimental}. Turbulence also introduces variations in the amplitude envelope of the acoustic pressure signal \citep{lieuwen2002experimental}, and such cycle-to-cycle variability reduces the time window for strong correlations. Thus, high values of cross-variable correlations will be sustained only for short epochs. 
\begin{figure}
    \centering
    \includegraphics[width=1\linewidth]{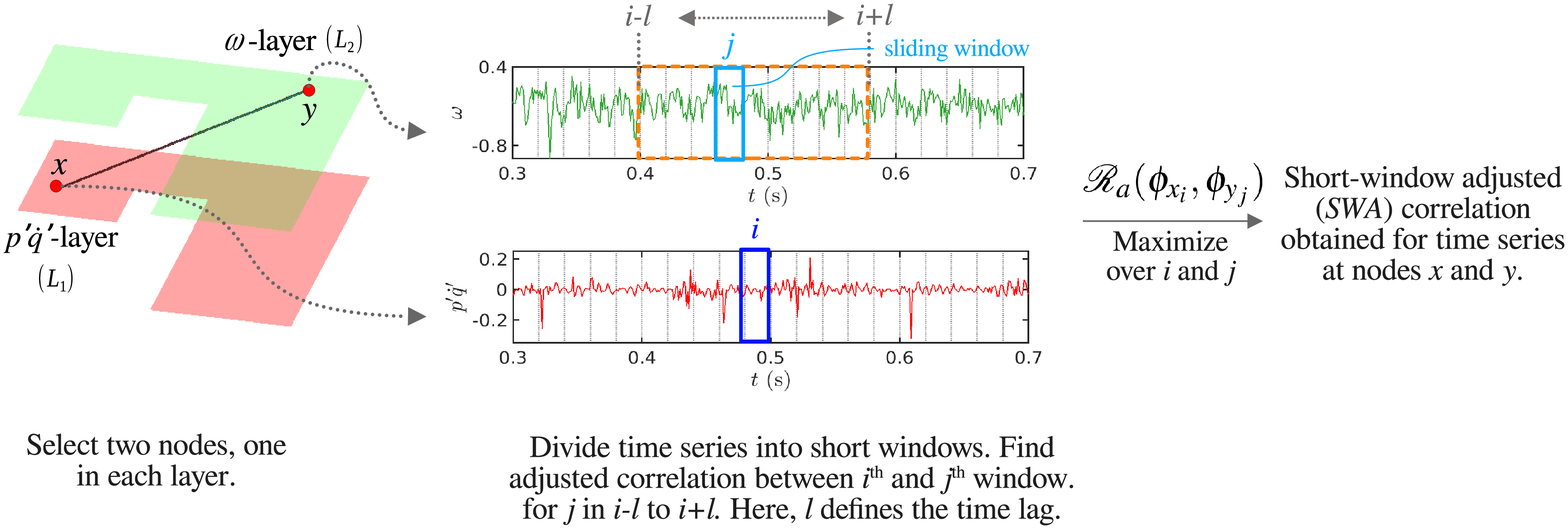}
    \caption{Flow chart depicting the procedure for determining short-window adjusted ($SWA$) correlation between time series at nodes $x$ and $y$ in layers $L_1$ ($\omega$-layer) and $L_2$ ($p'\dot{q} '$-layer), respectively.}
    \label{schem_ntwk_construct}
\end{figure}
\par Hence, we find cross-variable adjusted correlation for short epochs between the time series of $p'\dot{q} '$ and $\omega$ for different pairs of locations to determine the inter-layer connections. We call this correlation measure as short-window adjusted ($SWA$) correlation. Figure \ref{schem_ntwk_construct} summarizes the procedure of determining $SWA$-correlation measure. Consider two time series, $t_x$ and $t_y$ obtained at nodes $x$ and $y$ in layers $L_1$ and $L_2$ of the inter-layer network, respectively. We divide each time series into short non-overlapping windows, where each window has a size of $w$ seconds. This window is equivalent to $k_w$ integral number of cycles of the dominant mode of acoustic oscillations ($p'$) that occur at $f=140$ Hz frequency. Thus, $k_w$ cycles are equivalent to $w=k_w/f$ seconds. In Appendix \ref{App_sizewindow}, we discuss the effect of using different values of $k_w$ on the range of correlations. The same window size $w$ is used across all dynamical states. The data in the $i^\textrm{th}$ window of the time series $t_x$ is then written as $\phi_{x_i}=t_x[(i-1)w~:~i w]$. 

\par Note that, the two time series $t_x$ and $t_y$ correspond to different variables representing dynamics of different subsystems. Such variables not only ensue at different time scales, but also influence each other after a certain time delay, associated with the shedding and convection of coherent structures in the flow field or mixing and combustion processes \citep{lieuwen2005combustion, george2018pattern}. To incorporate such time delays, we calculate the correlation between the $i^{th}$ window of $t_x$ and a range of windows from $i-l$ to $i+l$ of the time series $t_y$, where, $l$ determines the maximum time-lag allowed for correlating two variables. Similarly, correlation is calculated between all windows of the two time series and the maximum (maximized over $i$ and $j$) of such correlation values is defined as the $SWA$-correlation. The definition of short-window adjusted correlation ($\rho$) is given by equation \ref{corr_eq1}. 
\begin{equation}\label{corr_eq1}
\rho(t_x,t_y) \coloneqq \max_{\substack{(i,j)\\ j \in [i-l,~ i+l] \\ i\in [1:N_w] }} \Big[ \big\lvert~\mathscr{R}_a (\phi_{x_i},\phi_{y_j}) ~\big\rvert \Big]
\end{equation}
where, $N_w$ is the number of windows a time series is divided into, and $i$ and $j$ are the window indices. Here, the optimum window size is chosen by finding the number of cycles ($k_w$) in a window for which the mean of $SWA$-correlations of time series of all pairs of nodes in distinct layers is maximum. We vary $k_w$ between three to ten and find that such optimum window size occurs when $k_w=4$. Here, $l=10$ corresponding to a maximum time-lag of $0.2$ s is chosen to allow us to capture delays associated with hydrodynamic convection timescales. Note that, the time delay associated with correlation between time series at any two spatial locations is variable and depends on the distance between the spatial points and their location with respect to the combustor walls or the bluff-body. The use of $SWA$-correlation helps extract the maximum of the short-window correlation between the two time series corresponding to such unknown variable time-lags for any pair of locations. In Appendix \ref{App_sizewindow}, we show that although the spatial mean of correlation values decreases as $k_w$ increases, the qualitative topology of the network derived remains the same.

\par In summary, we use short-window adjusted correlation $\rho_{xy}$ to determine the weights of inter-layer edges between nodes $x$ and $y$ in layers $L_1$ and $L_2$ respectively, thus accounting for time-lagged correlation between different spatial locations. The resulting $SWA$-correlation matrix is the adjacency matrix $A$ of the inter-layer network where, $A[x,y]= \rho_{xy}$. Note that, the adjacency matrix $A$ is not symmetric. Particularly, in our system, the inter-layer link with weight $A[x,y]$ gives the correlation between the time series of $p'\dot{q}'$ at node $x$ in the $p'\dot{q}'$-layer ($L_1$) and the time series of vorticity at node $y$ in the $\omega$-layer ($L_2$). On the other hand, $A[y,x]$ gives the correlation between the time series of vorticity at node $y$ in the $\omega$-layer and the time series of $p'\dot{q}'$ at node $x$ in the $p'\dot{q}'$-layer. Thus, $A[x,y]\neq A[y,x]$, and hence $A$ is an asymmetric matrix.
\section{Tools for multilayer network analysis \label{sec_tools_multilayer}}
\par In this section, we discuss the basic properties of multilayer networks that are used in the current work to analyse the network topology, namely, inter-layer node strengths, inter-layer network assortativity and inter-layer link-rank distribution. We discuss the utility of these tools in identifying the topology of the inter-layer network, and hence, the patterns of inter-subsystem interactions during different dynamical states in a thermoacoustic system. 
\subsection{Inter-layer node strength}
The inter-layer node strength of a node in a particular layer is defined as the sum of the weights of its inter-layer links \citep{bianconi2018multilayer}. In a two-layered network, a node $x$ in layer $L_1$ has a node strength defined by equation \ref{eq_NS_1}. The inter-layer node strength is normalised by the maximum possible inter-layer connections of a node.
\begin{equation}\label{eq_NS_1}
    NS_{x_{L_1}}=\Big(\sum_{j_{L_2}=1}^N A[x_{L_1},j_{L_2}]\Big)\big/N
\end{equation}
where $N$ is the total number of nodes in each layer and $A[x_{L_1},j_{L_2}]$ is the weight of the link between node $x$ in layer $L_1$ and node $j$ in layer $L_2$. Clearly, the node strength of node $x$ in layer $L_2$ (defined by equation \ref{eq_NS_2}) is different than that of node $x$ in layer $L_1$, since the adjacency matrix $A$ is not symmetric (refer Section \ref{sec_ntwkconstruct}). 
\begin{equation}\label{eq_NS_2}
    NS_{x_{L_2}}=\Big(\sum_{k_{L_1}=1}^N A[k_{L_1},x_{L_2}]\Big)\big/N
\end{equation}
\par The inter-layer node strength of a node in the $p'\dot{q}'$-layer signifies if the thermoacoustic power generated at that location is strongly correlated with the hydrodynamic activity at several other locations in the combustion chamber. Similarly, the inter-layer node strength of a node in the $\omega$-layer signifies if the vorticity dynamics at that location is correlated to the thermoacoustic power generated at different locations in the combustion chamber. Also, nodes with very high node strengths are referred to as hubs. The spatial distribution of such node strengths in the combustor geometry can help us identify hubs of the inter-layer network which are essentially spatial pockets having significant inter-subsystem influence. 

\par Also, along with such spatial distribution of node strengths, we must decipher the entire topology of inter-layer connections. For example, we are interested to know if there are any significant hub-to-hub connections across layers; or are the inter-layer connections of a hub in one layer spread over a large spatial region in the other layer. To do so, we borrow the concept of assortativity from single-layer networks and extend it to inter-layer networks.
\subsection{Inter-layer network assortativity}
\par Assortativity measures the tendency of a node to connect to other nodes with similar characteristics in the network \citep{barabasi2013networkbook}. Particularly, degree-assortativity examines if high-degree nodes tend to connect to similar high-degree nodes or not \citep{newman2002assortative, barabasi2013networkbook}. Degree is the total number of connections that a node has in a network. In single-layer networks, a standard measure of assortativity is the degree correlation function ($k_{nn}$) which captures the relation between the degree of connected nodes \citep{pastor2001dynamical, vazquez2002large, barabasi2013networkbook}. The degree correlation function $k_{nn}$ is defined by equation \ref{Eq_deg_corr}.
\begin{equation}\label{Eq_deg_corr}
    k_{nn}(k)=\sum_{k'} k' P(k'|k)
\end{equation}
where $P(k'|k)$ is the conditional probability of finding a node of degree $k'$ amongst the set of nodes that are connected to a node with degree $k$. All the nodes which are connected to a particular node are said to be its neighbors in the network. Note that, neighbors of a node in the network are not necessarily its spatial neighbors. Thus, $k_{nn}(k)$ quantifies the average degree of the neighbors of all those nodes which have degree $k$  \citep{barabasi2013networkbook}. In an assortative network, similar-degree nodes have high tendency to connect. As a result, the higher the degree of a node, the higher will be the average degree of its neighbors, and the degree correlation function $k_{nn}(k)$ is expected to increase with $k$ for an assortative network. On the other hand, the degree correlation function decreases with $k$ for disassortative networks. 

\par Degree correlations are essential to study the topology of inter-layer connections in a multilayer network \citep{reis2014avoiding}. Recently, \citet{de2016degree} proposed the extension of degree correlations for multilayer networks based on tensor notation and analysed the effect of assortativity on epidemic spreading in such networks. Similarly, we introduce here an extension of the definition of degree correlations for a two-layered network consisting of inter-layer connections alone. We represent the multilayer network by what is called a supra-adjacency matrix, which is a block matrix built from the adjacency matrices of the intra-layer and inter-layer networks \citep{bianconi2018multilayer}. The diagonal blocks of the supra-adjacency matrix represent the intra-layer connections and the off-diagonal blocks represent the inter-layer connections in the multilayer network. The diagonal blocks are void in our case since we consider the inter-subsystem connections alone in this study. The off-diagonal block entries are derived from the adjacency matrix $A$ as discussed below.

\par We wish to examine the distribution of the inter-layer links that represent strongly correlated inter-subsystem activity between two locations. To do so, we set a threshold $\rho_{th}$ on correlation values and obtain an unweighted matrix $A_u$ from the adjacency matrix $A$, such that if $A[x,y]<\rho_{th}$, then $A_u[x,y]=0$, else $A_u[x,y]=1$. We construct an unweighted supra-adjacency matrix $B$ from the inter-layer network using the unweighted matrix $A_u$, as defined in equation \ref{eq_supra_adj}.
\begin{equation}\label{eq_supra_adj}
B =    \begin{pmatrix}
0 & A_u\\
A_u^T & 0
\end{pmatrix}
\end{equation}

\par The supra-adjacency matrix therefore represents a network of $2N$ nodes with $N$ nodes in each layer. In our system, we note that the replica nodes in each layer essentially represent the same spatial point in the combustor. However, we treat them as distinct nodes in this formalism as they possess distinct information arising from different subsystems. Then, we calculate the degree correlation function $k_{nn}$ for the matrix $B$ using the conventional definition as given by equation \ref{Eq_deg_corr}. Here, $k_{nn}$ is referred to as the inter-layer degree correlation function. We study the variation of $k_{nn}$ with the inter-layer degree $k$ which is derived as the sum of number of inter-layer links of a node in the unweighted network represented by $B$. For the current analysis, we set $\rho_{th}=0.75$. The effect of varying the correlation threshold on the degree correlations is discussed in Appendix \ref{App_corrthreshold}.

\par Examining the degree correlations in an inter-layer network helps us infer the average tendency of a node with high (or low) degree in one layer to form connections with nodes having similar or dissimilar degrees in the other layer. For an inter-layer network, if $k_{nn}(k)$ increases with $k$, the inter-layer connectivity is assortative, i.e., a node with high degree in one layer tends to connect only to nodes with high degrees in the other layer on an average. For a thermoacoustic system, an assortative inter-layer network would imply that inter-subsystem activity is restricted to regions with nodes having high degrees. On the other hand, if $k_{nn}(k)$ decreases with $k$, the inter-layer network is disassortative and a hub in one layer tends to have many connections to nodes with low or moderately high degrees in the other layer. For a thermoacoustic system, a disassortative topology would imply that the inter-subsystem interactions is not restricted between hub-regions obtained in the two layers. Rather, activity in the hubs in one layer is correlated to the activity in a large spatial regime comprising the low-degree nodes in the other layer. Further, the variation of the degree correlation function can be approximated as a power law, as in equation \ref{eq_powerlawdegcorr} \citep{barabasi2013networkbook}.
\begin{equation}\label{eq_powerlawdegcorr}
    k_{nn}(k) \sim k^\mu 
\end{equation}
where $\mu$ is the correlation exponent. Real-world networks can exhibit different values of $\mu$ for different range of node degrees $k$. Such mixed assortativity has been observed earlier, for example, in citation networks \citep{barabasi2013networkbook}.  

We note that, usually there are only a few hubs in the network. These few hubs may have some connections amongst themselves. However, by definition, any given hub possesses a large number of connections, most of which are to nodes with low or moderately-high degrees. That is, a hub may have a few significant hub-to-hub connections and also large number of disassortative connections. As a result, the hub induces negative degree correlations \citep{small_pre_2009revising}. However, disassortativity does not rule out the existence of hub-to-hub connections which is a local tendency \citep{noldus2015assortativity}. The degree-correlation method, while useful to interpret the average tendency of connections of a node, cannot help infer the local hub-to-hub connections in a network \citep{noldus2015assortativity,litvak2013uncovering}. In a multilayer network, despite disassortative degree correlations, an inter-layer network can have significant hub-to-hub links between layers. It is therefore necessary to characterize the distribution of links between hub-regions and low-degree regions across the two layers to infer the extent of inter-subsystem activity in our combustor. Thus, we propose the use of link-rank distributions between nodes in different layers as described in the following sub-section.

\subsection{Inter-layer link-rank distribution}
\par The high-degree nodes in a network largely determine the topology of the network and the subsequent interpretations. These large degree nodes in a network are said to be `rich nodes'. If the rich nodes are interconnected, they form a group referred to as the rich-club \citep{zhou2004rich}. Despite strong disassortative degree correlations induced by hubs \citep{small_pre_2009revising}, networks can contain rich-clubs owing to local assortativity of hub-nodes \citep{zhou2004rich}. In an inter-layer network, a rich club would signify the presence of dense inter-layer connections between high-degree nodes across layers. Particularly, the existence of a rich-club in disassortative networks would signify that there are several connections amongst nodes with high degrees as well as significant connections between hubs and low-degree nodes. Thus, we must characterize the distribution of inter-layer links between high as well as low degree nodes in both the layers.

\par
\begin{figure}
    \centering
    \includegraphics[width=0.8\linewidth]{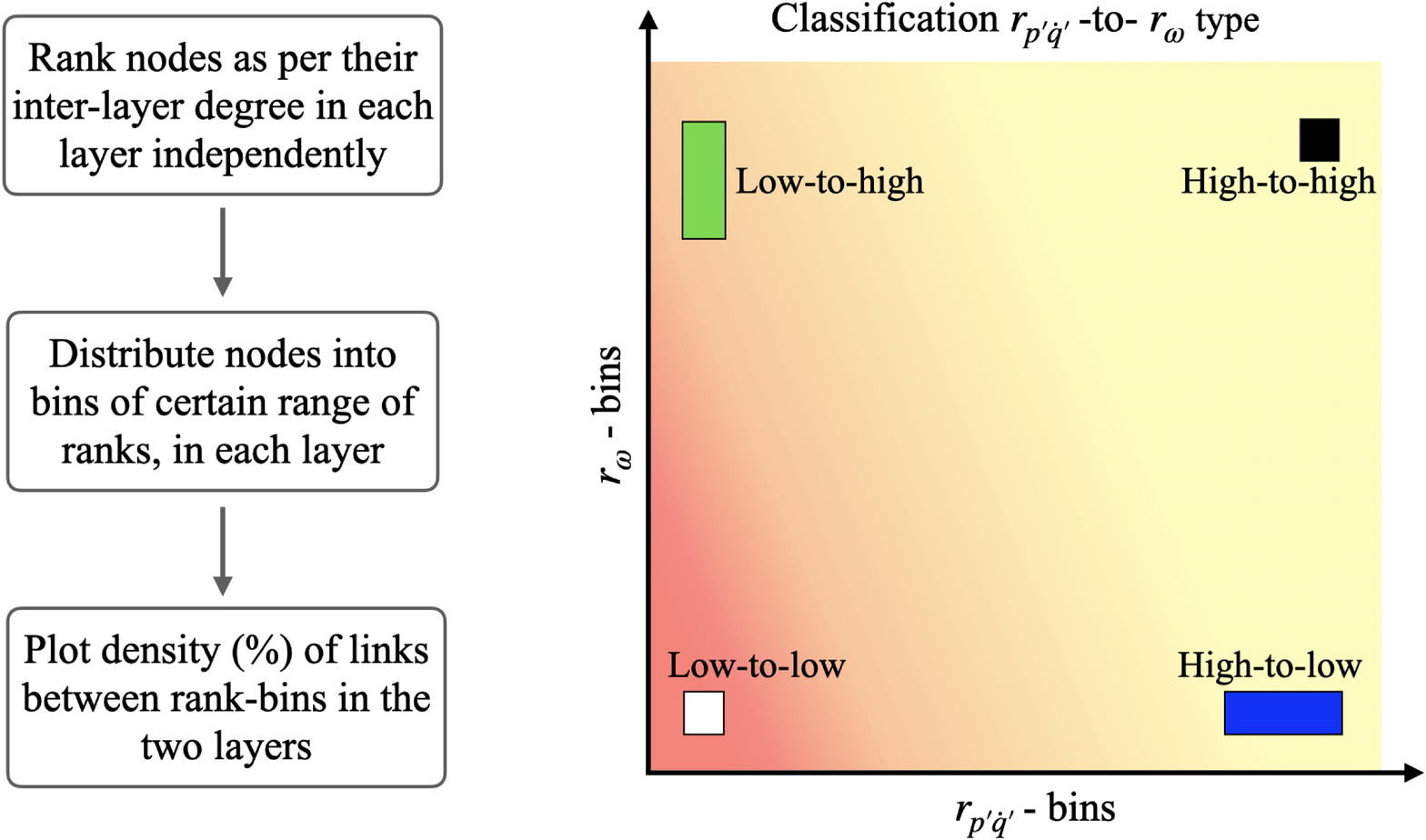}
    \caption{Schematic guideline for constructing and analysing inter-layer link-rank distributions and classifying different type of inter-layer links.}
    \label{schem_iln_linkrank}
\end{figure}

 To do so, we construct an inter-layer link-rank distribution as described in figure \ref{schem_iln_linkrank}. First, we rank the nodes in each layer according to their inter-layer degrees. The highest degree node in a layer has rank one and so on. Note that, the ranking of nodes in each layer is independent of the nodes in the other layer. We thus assign ranks $r_{p'\dot{q}'}$ to nodes in the $p'\dot{q}'$-layer and ranks $r_{\omega}$ to nodes in the $\omega$-layer. Next, we form bins of ranks in each layer where each bin classifies a range of ranks. For example, bin one constitutes nodes having the first 10\% ranks, the second bin constitutes nodes with the next 10\% ranks and so on. Thus, nodes in each layer are distributed into ten bins according to their ranks. Next, we examine what fraction of the total inter-layer links connect nodes in the $m^{th}$ bin of the $p'\dot{q}'$-layer and nodes in the $n^{th}$ bin of the $\omega$-layer (where $m$ and $n$ vary from one to ten). This gives us a distribution of inter-layer links classified according to the ranks (degrees) of the nodes connected by these links. We call such a distribution as an inter-layer link-rank distribution. Note that, this inter-layer link-rank distribution proposed here is inspired from the node-node link distribution suggested by \citet{zhou2004rich} for single-layer networks.
\par We can categorize the inter-layer links according to the ranks of the nodes they connect, as shown in figure \ref{schem_iln_linkrank}. Links that connect nodes with ranks $r_{p'\dot{q'}}$ in the ${p'\dot{q}'}$-layer and nodes with ranks $r_{\omega}$ in the $\omega$-layer are referred as $r_{p'\dot{q}'}-\text{to}-r_{\omega}$ link. If these links connect nodes in the low-ranking bins (high-degree nodes) in the ${p'\dot{q}'}$-layer to nodes in the high-ranking bins (low-degree nodes) in the $\omega$-layer, then such inter-layer links will be called low-to-high type links, as represented by a green box in the schematic inter-layer link-rank distribution in figure \ref{schem_iln_linkrank}. Similarly, there can be low-to-low (white box in figure \ref{schem_iln_linkrank}), high-to-low (blue box in figure \ref{schem_iln_linkrank}) and high-to-high (black box in figure \ref{schem_iln_linkrank}) type of links in an inter-layer network. 

\par The link-rank distribution is very helpful in deciphering several features of the inter-layer network topology. Firstly, if the percentage of low-to-low type of links is high in an inter-layer network, we can infer that there are dense hub-to-hub inter-layer connections in the network. Also, if the percentage of low-to-high links is very significant, we infer that hubs in the first layer connect to regions of low node strengths in the other layer. Secondly, we can also compare the different types of links to infer the density of inter-layer connections between different regions of the combustor. 

We note that, the inter-layer degree correlation function ($k_{nn}$) captures the average type of connections each node tends to have. On the other hand, link-rank distribution is obtained by binning several similar degree nodes together and shows the fraction of connections between these similar-degree nodes. Thus, link-rank distribution reflects the number of connections a group of nodes have. Using both inter-layer degree correlation function and inter-layer link-rank distribution, we examine the topology of the unweighted inter-layer network obtained by setting the correlation threshold $\rho_{th}=0.75$ during various dynamical states in a turbulent combustor and the results are presented in Section \ref{sec_results}. 
\section{Results\label{sec_results}}
\par We use two-layered inter-layer networks to analyse the inter-subsystem interactions during diverse dynamical states in a turbulent bluff-body stabilized combustor. To assess the raw flow field, we plot the normalised time series of acoustic pressure ($p'$) and heat release rate fluctuations ($\dot{q}'$) (row-I) along with the spatio-temporal evolution of the $p'\dot{q}'$ field (row-II) and vorticity field (row-III) in figures \ref{fig_raw_480}, \ref{fig_raw_570_aperiodic}, \ref{fig_raw_570_periodic}, \ref{fig_raw_765} during the occurrence of combustion noise (chaos), aperiodic and periodic epochs of intermittency and thermoacoustic instability (order), respectively. The $p'$ and $\dot{q}'$ time series are normalized with the corresponding maximum value during each dynamical state to facilitate comparison across different dynamical states. Next, using the tools defined in Section \ref{sec_tools_multilayer}, we analyse how the topology of the inter-layer connections changes as the dynamics of the combustor transitions from chaos to order via the route of intermittency (figures \ref{fig_480_ILN}, \ref{fig_570_ILN}, \ref{fig_765_ILN}). To aid in understanding the topology of the inter-layer network, we also plot a multilayer visualization of the inter-layer connections for specific nodes in both the layers (figures \ref{fig_480_visual_ILN}, \ref{fig_570_visual_ILN}, \ref{fig_765_visual_ILN}). We consider one sample node in the recirculation zone in the dump plane upstream of the bluff-body and another in the wake downstream of the bluff-body. The two-layered visualization is derived from the inter-layer network constructed from the experimental data during distinct dynamical states.
\subsection{Combustion noise}
\par
\begin{figure}
    \centering
    \includegraphics[width=1\linewidth]{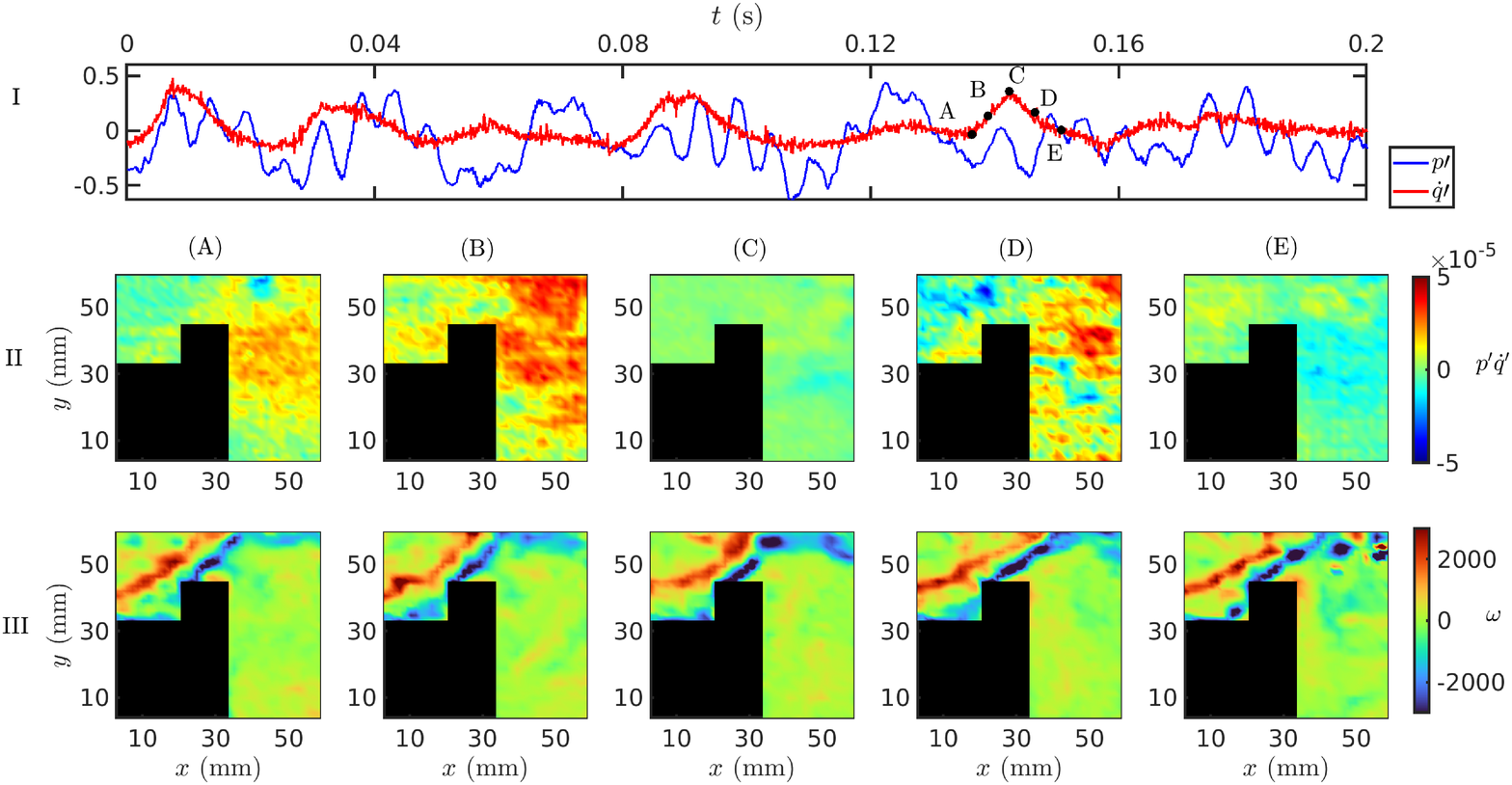}
    \caption{(I)-Time series of acoustic pressure ($p'$) and global heat release rate fluctuations ($\dot{q}'$) during the state of \textbf{combustion noise}. Spatial distribution of (II)-thermoacoustic power generation and (III)-vorticity at time instants corresponding to points A-D shown in the time series in (I). Thermoacoustic power sources exist predominantly in the wake region and incoherent patterns of $p'\dot{q} '$ are observed. Small-sized vortices are shed at irregular epochs in the recirculation zone and in the wake due to which the vorticity is high in the shear layers.}
    \label{fig_raw_480}
\end{figure}

 During the state of combustion noise, the combustor exhibits low-amplitude chaotic fluctuations in the acoustic pressure ($p'$) and global heat release rate ($\dot{q}'$) signals, as shown in figure \ref{fig_raw_480}-(I). Further, the value of $p'\dot{q}'$ is small in magnitude and we observe incoherent spatial patterns in the $p'\dot{q}'$ field observed throughout the turbulent reacting flow field of the combustion chamber, as evident in figure \ref{fig_raw_480}-(II) \citep{george2018pattern}. Also, patches of slightly high values of $p'\dot{q}'$ occur in the wake of the bluff-body erratically as is evident from figure \ref{fig_raw_480}-(II)B.
\par In the vorticity field, we observe from experiments that small-sized vortices are shed in the recirculation zone in the dump plane and in the wake of the bluff-body. The vorticity is found to be high and positive in the outer shear layer (figure \ref{fig_raw_480}-(III)) owing to the anticlockwise vortices shed in this region. The vortices shed in the wake are formed due to the flow turning at the tip of the bluff-body from where we see the origin of the inner shear layer. Vorticity is high and negative in the inner shear layer owing to the clockwise vortices are shed in the wake. The streaks of high vorticity in the inner and outer shear layer appear consistent in time, while the shedding of the small vortices is rather erratic \citep{george2018pattern}. 

\begin{figure}
    \centering
    \includegraphics[width=1\linewidth]{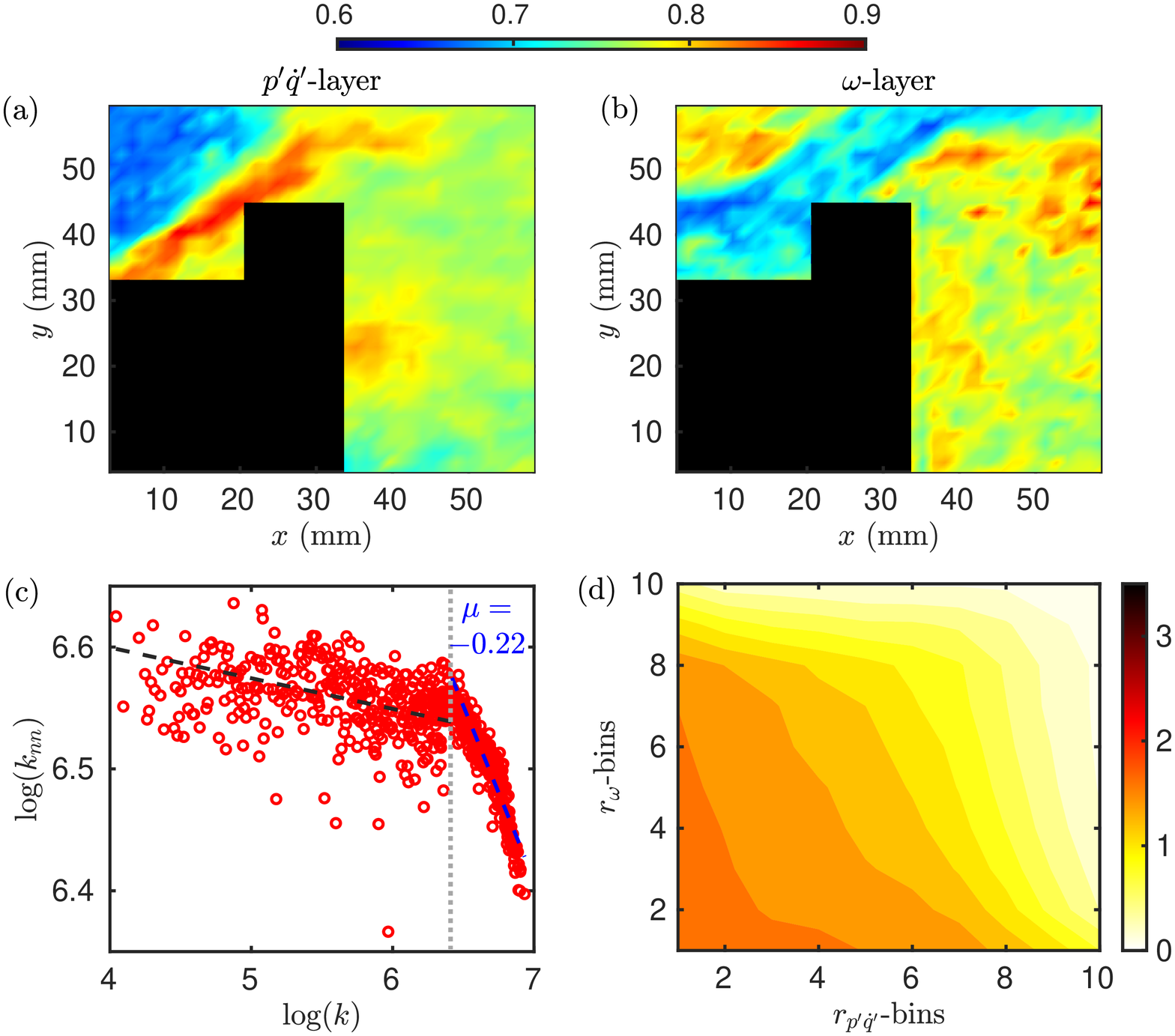}
    \caption{The spatial distribution of node strengths in (a) the $p'\dot{q} '$-layer and (b) the $\omega$-layer derived from the inter-layer network during the state of \textbf{combustion noise}. (c) Variation of the inter-layer degree correlation function ($k_{nn}$) with inter-layer degree ($k$). $k_{nn}$ exhibits a negative power law variation with $k$. For low-degree nodes, the network exhibits weak disassortativity, while for high-degree nodes the power law exponent of degree correlations is $\mu=-0.22$ implying strong disassortativity. (d) The inter-layer link-rank distribution of links between nodes ranked into bins in the $p'\dot{q}'$-layer and the $\omega$-layer. High density of inter-layer links exist between low ranking nodes in both layers. Also, significant links exist between low and high ranking nodes in both layers leading to disassortativity in the network. 
    }
    \label{fig_480_ILN}
\end{figure}
\par Results of the analysis of the inter-layer network constructed from the vorticity and the thermoacoustic power field during the occurrence of combustion noise are shown in figure \ref{fig_480_ILN}. We observe that the inter-layer node strengths in the $p'\dot{q}' $-layer are high in the shear layer region and moderately high in the wake region, but low in the recirculation zone in the dump plane (figure \ref{fig_480_ILN}(a)) during the state of combustion noise. In contrast, in the $\omega$-layer (figure \ref{fig_480_ILN}(b)), high inter-layer node strength is obtained in the recirculation zone in the dump plane and in the wake region, and very low node strength in the shear layers. 

\par Next, we study the topology of the inter-layer network obtained during the state of combustion noise (as explained in Section \ref{sec_tools_multilayer}). Figure \ref{fig_480_ILN}(c) shows that the variation of inter-layer degree correlation function ($k_{nn}$) with the inter-layer degree ($k$) exhibits distinct power-law behaviors for low and high range of degrees, i.e., the network exhibits mixed assortativity. In the low-degree range, $k_{nn}$ decreases with $k$ on an average, and the correlation exponent ($\mu$) is close to zero indicating neutral network topology with weak disassortativity. In the high-degree range, we observe strong disassortativity with $\mu=-0.22$. Figure \ref{fig_480_ILN}(d) shows the inter-layer link-rank distribution derived from the network during the state of combustion noise. We obtain high density of links between the low-ranked bins (low-to-low $r_{p'\dot{q}'}-\text{to}-r_{\omega}$ type). Also, significant inter-layer links exist between nodes in the two layers with moderate and high rankings. That is, the density of inter-layer connections of all types (low-to-low, high-to-low, low-to-high and high-to-high) are significant and comparable. 

\begin{figure}
    \centering
    \includegraphics[width=0.9\linewidth]{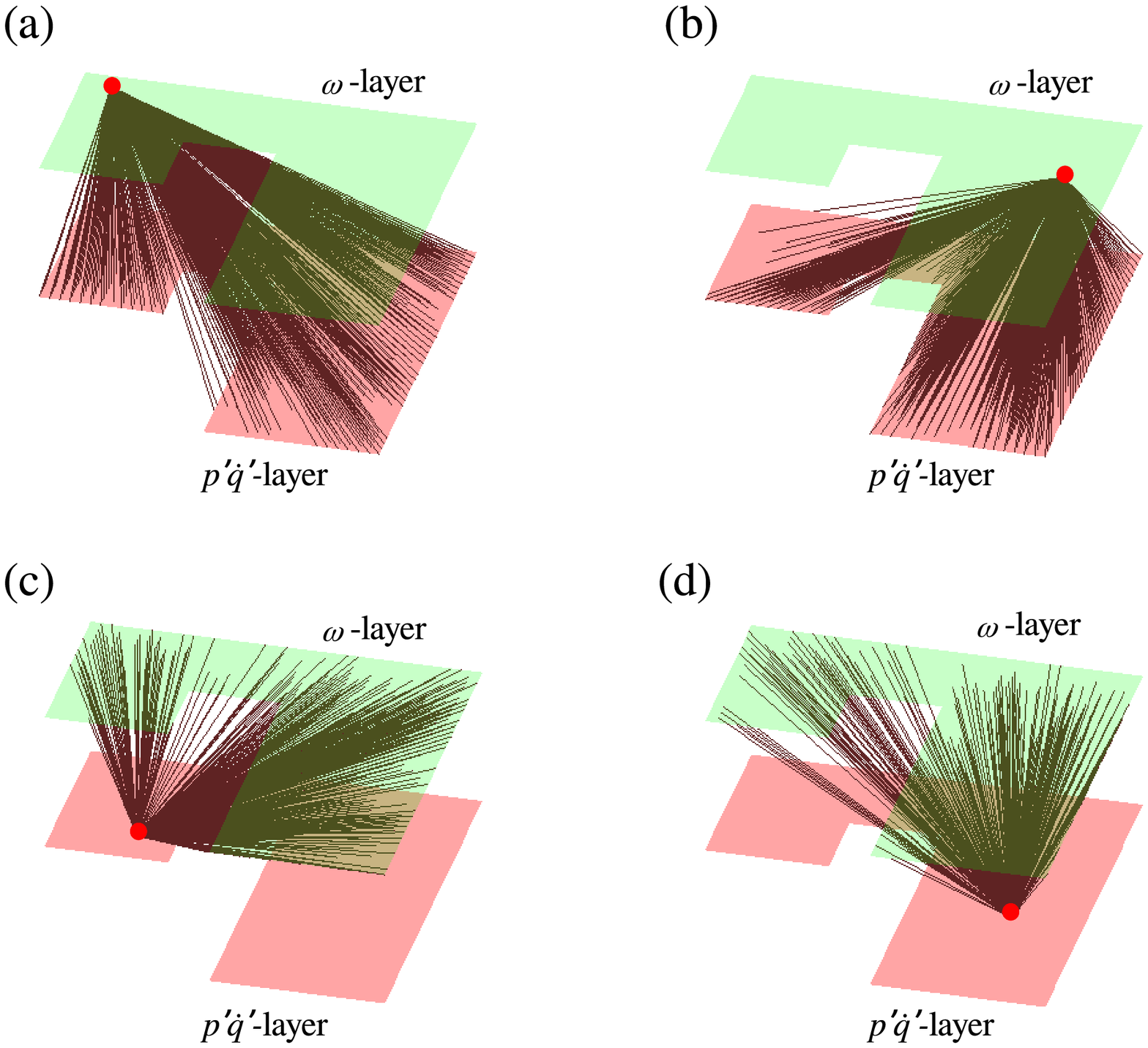}
    \caption{Visualization of inter-layer links derived from the inter-layer network during the state of \textbf{combustion noise} for sample nodes. The green surface represents the $\omega$-layer, while the red surface represents the $p'\dot{q}'$-layer. The cut out region in each layer indicates the location of the bluff-body. Sample nodes are selected (a) in the dump plane in the $\omega$-layer, (b) in the wake region in the $\omega$-layer, (c) in the dump plane in the $p'\dot{q}'$-layer, and (d) a node in the wake region in the $p'\dot{q}'$-layer. The inter-layer connections of most nodes are spread over wide spatial locations indicating non-localized inter-subsystem interactions in the combustion chamber.}
    \label{fig_480_visual_ILN}
\end{figure}

From the spatial distribution of node strengths we note that high-degree nodes are found in multiple locations in the spatially extended flow field, such as, in the outer shear layer in the $p'\dot{q}'$-layer or in the wake in the $\omega$-layer. The variation of the degree correlation function implies that on an average, a high-degree node has more disassortative connections. Further the link-rank distribution implies that these disassortative links arising from high-degree nodes connect to regions with high as well as low degree nodes that are spread throughout different locations in the combustor. Thus, we understand that inter-layer connections exist throughout the flow field and are not localised during the state of combustion noise. This interpretation is supported by the visualization of inter-layer links of sample points as shown in figure \ref{fig_480_visual_ILN}. 

Figure \ref{fig_480_visual_ILN}(a) shows that the inter-layer links of a node selected in the dump plane of the $\omega$-layer (green-colored layer) connect it to most nodes in the $p'\dot{q}'$-layer (red-colored layer) throughout the flow field. Figure \ref{fig_480_visual_ILN}(b) shows that a node in the wake region in the $\omega$-layer connects predominantly to nodes in the $p'\dot{q}'$-layer in the bluff-body wake and in the shear layer regions upstream of the bluff-body (regions of high node strengths in the $p'\dot{q}'$-layer, see figure \ref{fig_480_visual_ILN}(a)). Similarly, the inter-layer links of a node in the $p'\dot{q}'$-layer selected in the recirculation zone (\ref{fig_480_visual_ILN}(c)) are spread across in the $\omega$-layer in the recirculation zone and the wake, but not in the shear layer regions as expected from the spatial distribution of node strengths in figure \ref{fig_480_ILN}(b). Also, a node in the wake in the $p'\dot{q}'$-layer (see figure \ref{fig_480_visual_ILN}(b)) connects predominantly to nodes in the wake in the $\omega$-layer, with a few links spread into the recirculation zone. Clearly, the hydrodynamic activity and thermoacoustic power generated at nodes in the wake region are densely inter-connected. Also, inter-subsystem interactions occur between spatial locations upstream and downstream of the bluff-body.

We note that, $p'\dot{q}'$ fluctuations are low in the recirculation regions as evident in figure \ref{fig_raw_480}-II. Also as discussed by \citet{poinsot1987vortex} earlier, during the state of combustion noise, no reaction occurs in the recirculation zones where the vortices are shed. Here, we find low node strengths in the recirculation zone in the dump plane in the $p'\dot{q}'$-layer, indicating that thermoacoustic power generated in this region is small and weakly correlated to the vortex dynamics in the flow field. Further, low-node strengths are obtained in the shear layer in the $\omega$-layer owing to the incoherence induced in the vorticity field by erratic vortex shedding during the state of combustion noise. 

Clearly, while thermoacoustic power generation is maximum in the wake downstream of the bluff-body (figure \ref{fig_raw_480}-(II)), a region of high inter-layer node strength appears in the shear layer of the $p'\dot{q}'$-layer (figure \ref{fig_480_ILN}(a)), that is upstream of the bluff-body. Inter-subsystem interactions occur between spatially separated regions upstream and downstream of the bluff-body during the state of chaotic dynamics owing to the time delays associated with convection of vortices and fine-scale mixing of fuel and air. Therefore, from the network topology we infer that the thermoacoustic power generated in the wake of the bluff-body as well as the shear layer regions is highly correlated with the vortex dynamics in the recirculation regions in the dump plane and the wake. We infer that the small vortices that are shed in the dump plane convect downstream while the combustion of the reactants within these vortices is still in progress. As a result, the vorticity dynamics in the dump plane is correlated with the thermoacoustic power generated downstream, in the shear layer and wake of the bluff-body. Further, the vortices shed in the wake region also contain some unburnt reactant mixture which undergoes combustion in the wake leading to large thermoacoustic power generation in this region.

\subsection{Intermittency}
\par
\begin{figure}
    \centering
    \includegraphics[width=1\linewidth]{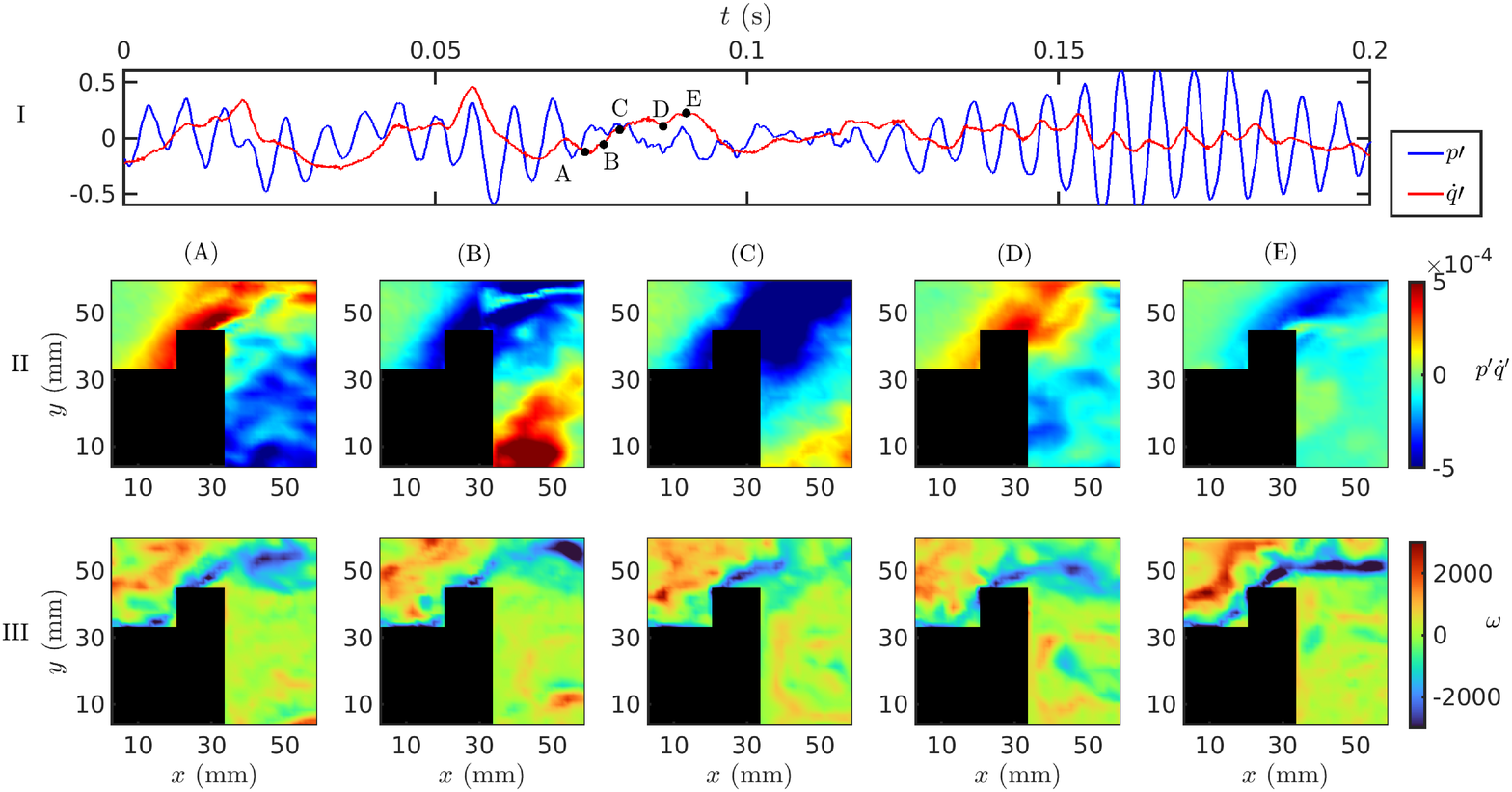}
    \caption{(I)-Time series of acoustic pressure ($p'$) and global heat release rate fluctuations ($\dot{q}'$) during an \textbf{aperiodic epoch of intermittency}. Spatial distribution of (II)-thermoacoustic power generation and (III)-vorticity at time instants corresponding to points A-D shown in the time series in (I). Thermoacoustic power source appears erratically while thermoacoustic power sink exists predominantly in the shear layer region. Positive incoherent vorticity field exists in the dump plane and the outer shear layer, while a negative vorticity streak is observed in the inner shear layer.}
    \label{fig_raw_570_aperiodic}
\end{figure}

\begin{figure}
    \centering
    \includegraphics[width=1\linewidth]{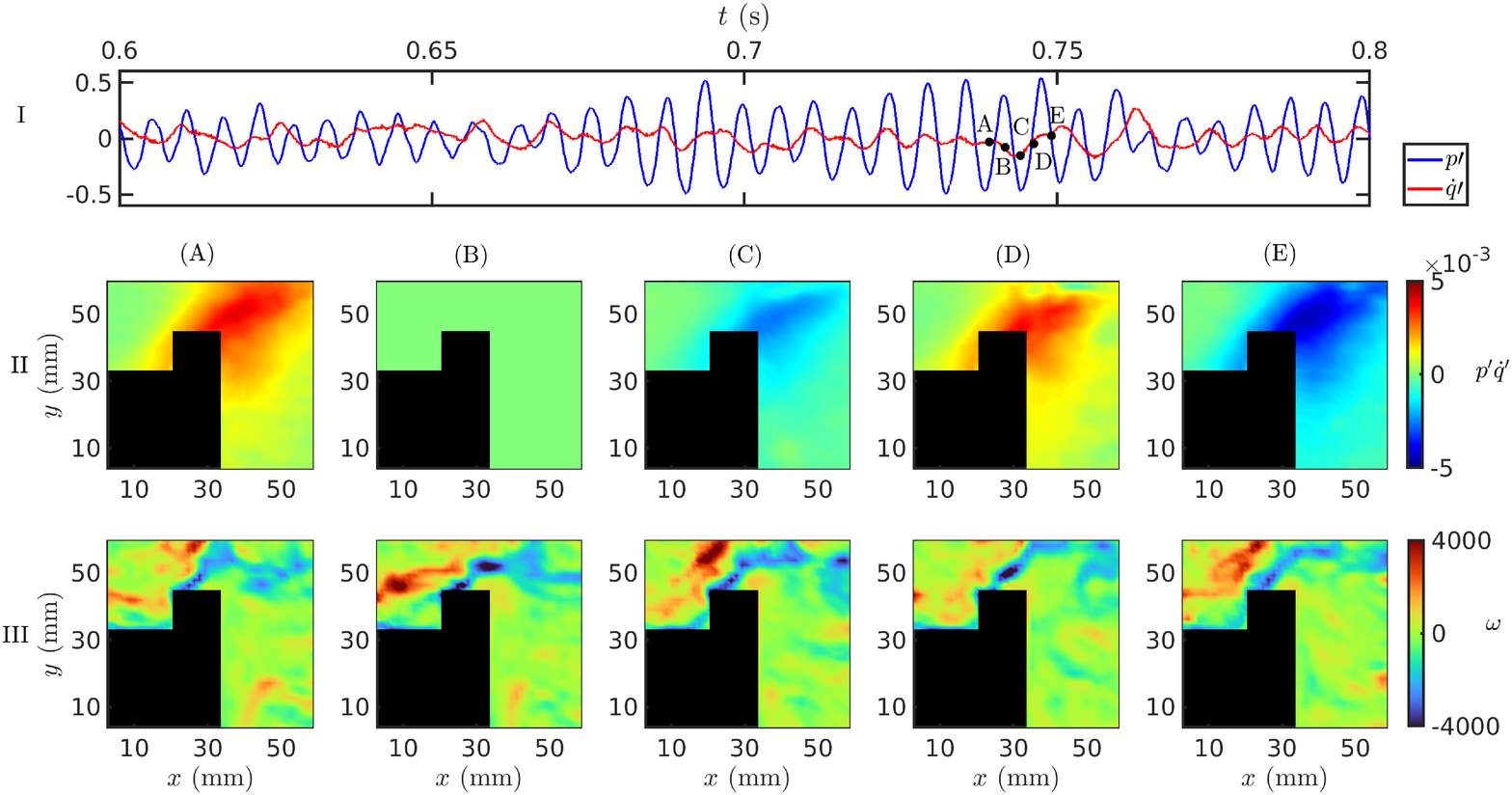}
    \caption{(I)-Time series of acoustic pressure ($p'$) and global heat release rate fluctuations ($\dot{q}'$) during \textbf{periodic epoch of intermittency}. Spatial distribution of (II)-thermoacoustic power generation and (III)-vorticity at time instants corresponding to points A-D shown in the time series in (I). Thermoacoustic power generated during periodic epochs is an order of magnitude higher than during the aperiodic epochs. Source of $p'\dot{q}'$ exists predominantly in the shear layer and also in the wake close to the bluff-body. Large patches of high positive vorticity appear and disappear in the dump plane and the outer shear layer regions due to the shedding of large coherent structures in these regions.}
    \label{fig_raw_570_periodic}
\end{figure}

 During the state of intermittency, we observe bursts of periodic oscillations amidst aperiodic fluctuations in the acoustic pressure and the heat release rate signals \citep{nair2014intermittency}. In figure \ref{fig_raw_570_aperiodic} and figure \ref{fig_raw_570_periodic}, we show the variations in the $p'\dot{q}'$ (row-II) and vorticity fields (row-III) in the combustion chamber during epochs of aperiodic and periodic acoustic pressure dynamics, respectively.

\par During aperiodic dynamics in intermittency, we observe sinks of thermoacoustic power generation (negative values of $p'\dot{q}'$ are predominant) as evident from figure \ref{fig_raw_570_aperiodic}-(II). Note that, the $p'\dot{q}'$ field obtained for aperiodic dynamics during the state of intermittency is very different from that during combustion noise, since we do not obtain incoherent patterns of $p'\dot{q}'$. Instead, we observe coherent sinks of $p'\dot{q} '$ with few erratic occurrences of $p'\dot{q}'$ sources predominantly in the shear layer region such as in figure \ref{fig_raw_570_aperiodic}-(II)A,D. On the other hand, during the periodic epochs of intermittency (see figure \ref{fig_raw_570_periodic}-(II)), we observe coherent regions that are source (or sink) of thermoacoustic power generation with values which are one order of magnitude higher than that during the aperiodic epochs. These regions of coherent $p'\dot{q}'$ generation occur in the inner shear layer and in the wake close to the bluff-body as observed in figure \ref{fig_raw_570_periodic}-(II). Such coherent source and sinks of $p'\dot{q}'$ occur periodically during the epochs of periodic dynamics in the state of intermittency. 

\par Further, from experiments we note that small vortices are shed during aperiodic dynamics in the state of intermittency. From figure \ref{fig_raw_570_aperiodic}-(III), we observe incoherence in the vorticity field fluctuations specifically in the recirculation zone. Unlike the vorticity field during the state of combustion noise, a consistent streak of vorticity does not exist in the outer shear layer. On the other hand, during periodic epochs of intermittency (see figure \ref{fig_raw_570_periodic}-(III)), we observe that the vortices formed in the recirculation zone are larger than those formed during the state of combustion noise or aperiodic epochs of intermittency. These large vortices are shed almost periodically for short epochs in the recirculation zone, after which they traverse to the wake via the shear layer and break down into smaller vortices or impinge on the walls of the combustor and the bluff-body. Thus, we find distinct patches of large positive vorticity appear and disappear in the recirculation zone (figure \ref{fig_raw_570_periodic}-(III)B) and then in the outer shear layer (figure \ref{fig_raw_570_periodic}-(III)C). In contrast, the vortices shed at the tip of the bluff-body are small. So, we obtain an almost consistent streak of negative vorticity in the inner shear layer (figure \ref{fig_raw_570_periodic}-(III)A-D).

\par
\begin{figure}
    \centering
    \includegraphics[width=1\linewidth]{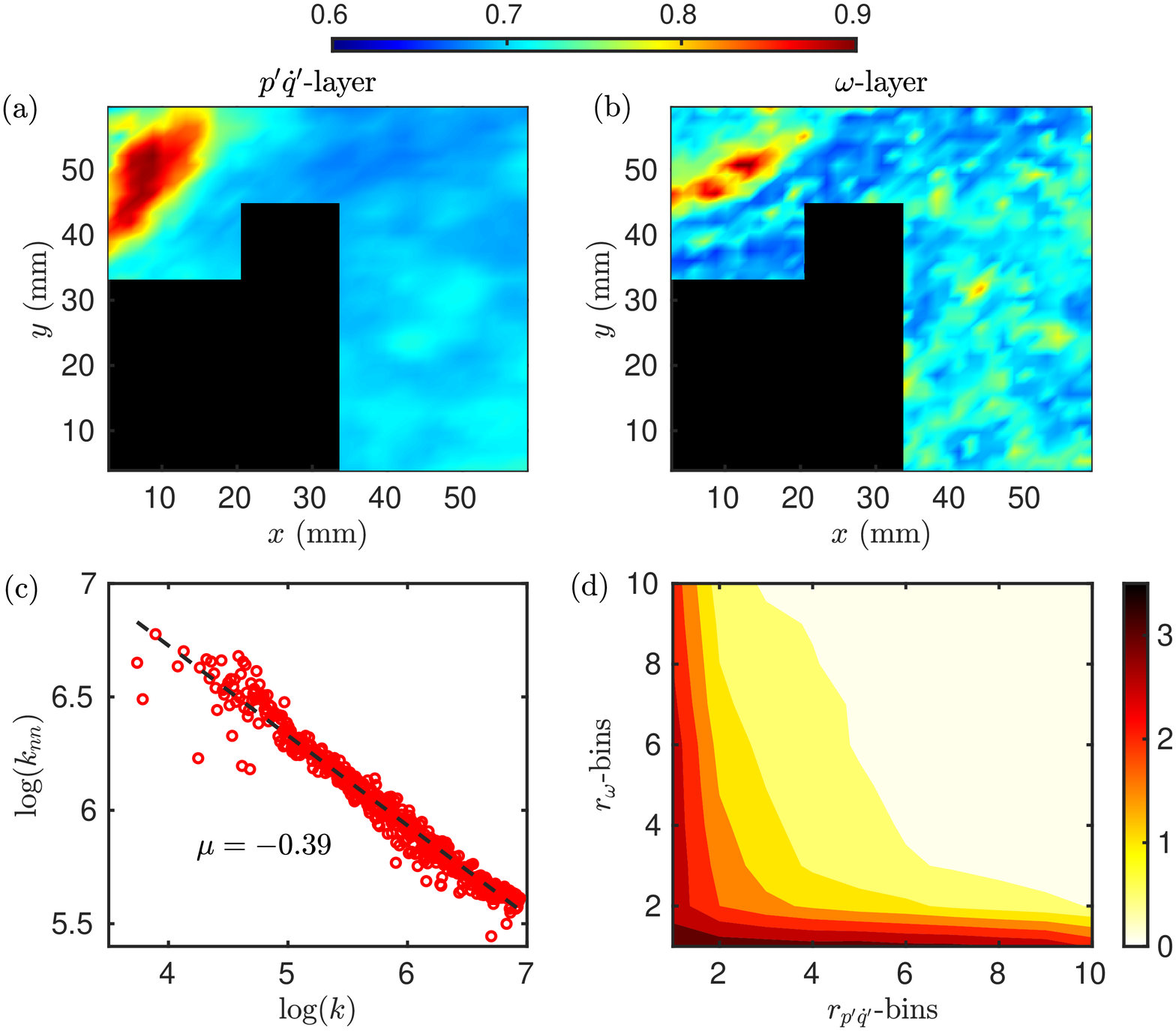}
    \caption{The spatial distribution of node strengths in the (a) $p'\dot{q} '$-layer and (b) $\omega$-layer derived from the inter-layer network during the state of \textbf{intermittency}. A spatial patch of very high node strengths (hubs) appear in the dump plane in both the layers. (c) Variation of the inter-layer degree correlation function ($k_{nn}$) with inter-layer degree ($k$), where the power law exponent is $\mu=-0.39$ indicating strong disassortativity. (d) The inter-layer link-rank distribution of links between nodes ranked into bins in the $p'\dot{q}'$-layer and the $\omega$-layer. The colorbar indicates the percentage of links. The density of links connecting hubs in both layers (low-to-low type links) is very high. Also, low-to-high and high-to-low type of links are significant indicating that hubs in one region connect also to regions of moderate or low node strengths in the other layer.}
    \label{fig_570_ILN}
\end{figure}

 Figure \ref{fig_570_ILN} shows the results from the inter-layer network analysis of the spatio-temporal dynamics occurring during both the periodic and aperiodic epochs of intermittency. The node strength distribution of both the layers (figure \ref{fig_570_ILN}(a,b)) is peculiar showing a distinct patch of very high node strengths (about 0.9) in the recirculation zone in the dump plane downstream of the backward-facing step (refer schematic figure \ref{fig_schematic}(c)); while in other regions the node strengths are moderately high (about 0.6). These localized pockets formed by a set of hub-nodes occur in similar regions in the dump plane in both the layers. 
\par The inter-layer degree correlation function ($k_{nn}$) derived from the inter-layer network decreases with increase in the inter-layer degree ($k$) and exhibits a power-law with correlation exponent $\mu=-0.39$, implying strong disassortativity in the topology of the network (figure \ref{fig_570_ILN}(c)). Such disassortative topology implies that, on an average a hub-node (in the dump plane) in one layer has several connections to nodes with low node strengths (in the shear layer and wake) in the other layer. Figure \ref{fig_570_ILN}(d) shows the inter-layer link-rank distribution. We find that a large fraction of inter-layer connections are of the low-to-low type, that is, several links exist between the high degree nodes (hubs) in both layers. Despite the fact that each hub node has a large number of disassortative links, we find that the spatial collection of such nodes concentrated in a localized pocket in each layer has significant number of hub-to-hub inter-layer links. Further, we find that high-to-low and low-to-high type of links have significant density, where as, very few connections are of the high-to-high type (figure \ref{fig_570_ILN}(d)). Both high-to-low and low-to-high type of links are equally significant; thus, we understand that the hubs in both the layers are equally influential. In other words, both, the thermoacoustic power generated and the vortices shed in the dump plane are equally crucial in fostering inter-subsystem interactions in the combustor. 

We note that, both, disassortativity and the existence of links between groups of low and high ranking nodes across layers together help us infer the network topology. In summary, during the state of intermittency, we obtain an inter-layer network with localised pockets of hub-nodes having dense hub-to-hub connections as well as disassortative connections throughout the flow field. The inter-layer network topology dictates that the thermoacoustic power generated in the recirculation zone in the dump plane is strongly correlated with the vorticity dynamics in all regions of the combustor during the state of intermittency. Also, the shedding of vortices in the recirculation zone of the dump plane is correlated over short epochs with the thermoacoustic power generated in all regions of the combustor.

\par
\begin{figure}
    \centering
    \includegraphics[width=0.9\linewidth]{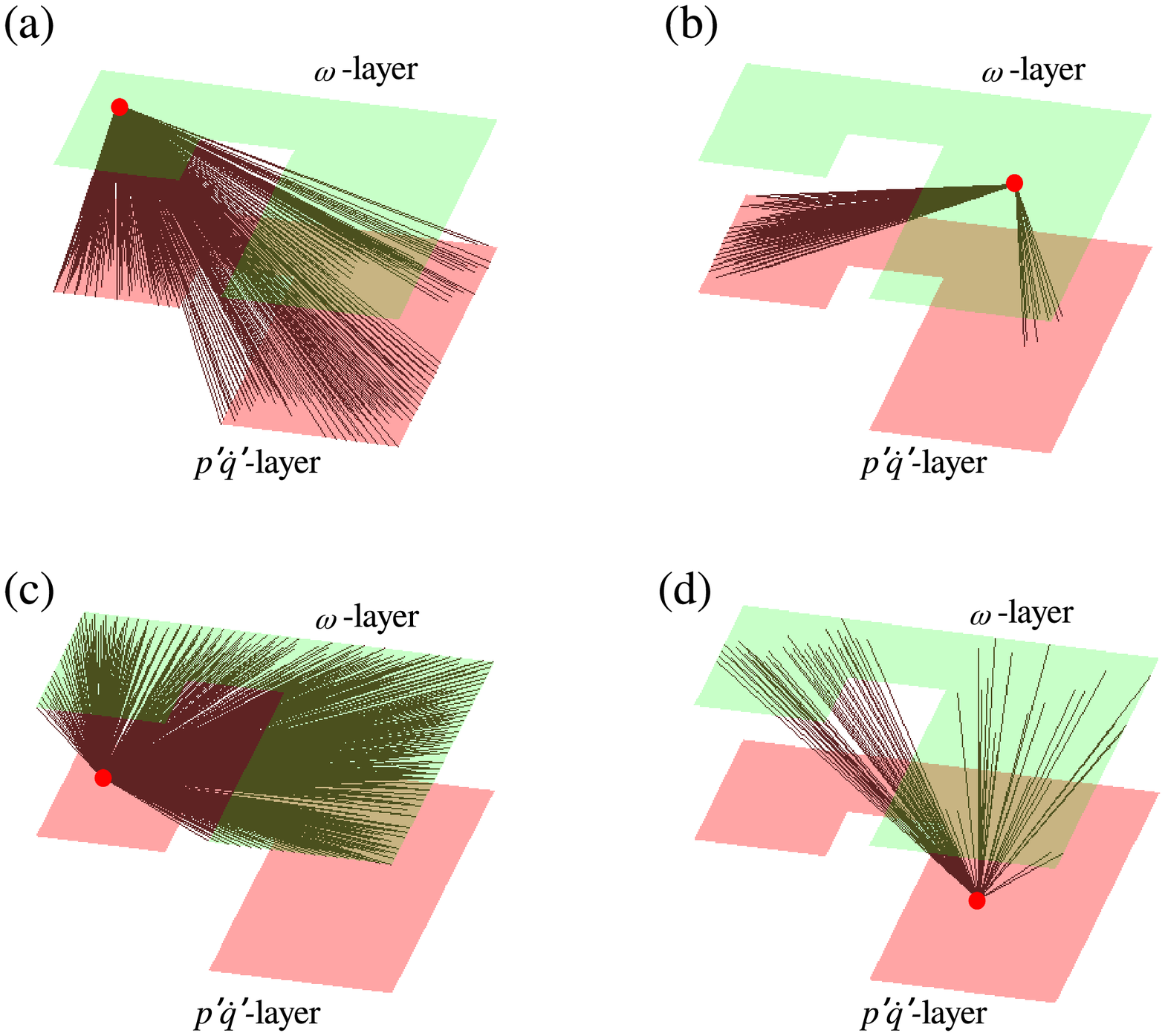}
    \caption{Visualization of inter-layer links derived from the inter-layer network during the state of \textbf{intermittency} for sample nodes. The green surface represents the $\omega$-layer, while the red surface represents the $p'\dot{q}'$-layer. The cut out region in each layer indicates the location of the bluff-body. Sample nodes are selected (a) in the recirculation zone in the $\omega$-layer, (b) in the wake region in the $\omega$-layer, (c) in the recirculation zone in the $p'\dot{q}'$-layer, and (d) a node in the wake region in the $p'\dot{q}'$-layer. The hub-region in each layer has dense connections to the hub-region as well as regions of low and moderate node strengths in the other layer, indicating the widespread influence of hubs on inter-subsystem activity during the state of intermittency.}
    \label{fig_570_visual_ILN}
\end{figure}

 A visualization of the inter-layer links of sample nodes in the two layers is shown in figure \ref{fig_570_visual_ILN}. A node in the recirculation zone in the hub-region of the $\omega$-layer (see figure \ref{fig_570_visual_ILN}(a)) has dense connections to nodes in the dump plane (hubs) of the $p'\dot{q}'$-layer as well as to many nodes in the wake region. Similarly, a node in the hub-region of the $p'\dot{q}'$-layer (see figure \ref{fig_570_visual_ILN}(c)) connects to nodes in all regions, namely, the dump plane, the shear layer and the wake of the bluff-body. On the other hand, a node selected in the wake region in either layer has significant inter-layer connections only to the hub region in the other layer and with some nodes in its neighboring spatial locations in the wake (see figure \ref{fig_570_visual_ILN}(b,d)). 

\par Remember, large vortices are shed in the recirculation zone in the dump plane, while small vortices are shed in the wake of the bluff-body (figure \ref{fig_raw_570_periodic}-(III)) during the periodic epochs of intermittency. These large vortices induce higher local strain rates in regions between vortices \citep{schadow1992combustion}, and thus enhancing the mixing of the fuel-air mixture in some of these regions, thus promoting combustion of fuel-air mixture contained in these vortices. However, larger vortices also promote bulk mixing and can delay molecular mixing in some other regions, as a result of which some unburnt reactant mixture is convected downstream. We conjecture that, during the state of intermittency, some of the fuel-air mixture carried by the larger vortices that are shed in the recirculation zone undergoes combustion in the recirculation zone itself, specifically during the periodic epochs of intermittency. As a result, the vorticity dynamics and the thermoacoustic power generated in these spatial pockets (identified as hubs of the network) are strongly correlated, and we obtain dense hub-to-hub connections in the inter-layer network. 

Further, the disassortative connections between the nodes in the $p'\dot{q}'$ and $\omega$ layers indicate that the vorticity and $p'\dot{q}'$-fields in the dump plane are strongly correlated to the vorticity dynamics and thermoacoustic power generated downstream. Such correlations between upstream and downstream locations are similar to those obtained during the state of combustion noise. Thus, we conjecture that, the smaller vortices shed during the aperiodic epochs of intermittency, convect the reactant mixture downstream and cause thermoacoustic power generation in these locations (figure \ref{fig_raw_570_aperiodic}-II). Also, some unburnt fuel-air mixture may be retained from the large vortices in the dump plane and is convected downstream while it undergoes combustion. As a result, intense and coherent heat release occurs not only in the dump plane but also in the subsequent downstream locations (as evident in figure \ref{fig_raw_570_periodic}-(II)A,D), that is, in the outer and inner shear layers.

\par In summary, using multilayer networks, we identify localized hub-regions in the recirculation zone behind the backward-facing step (i.e., the dump plane) where intense inter-subsystem interactions emerge during the state of intermittency. Note that, such significant inter-subsystem interactions were believed to occur only during thermoacoustic instability when order is established in the spatio-temporal dynamics of the combustor \citep{poinsot1987vortex, schadow1992combustion}. However, we show using multilayer network analysis that such interactions emerge much prior to the onset of order, during the state of intermittency. Further, the framework of multilayer networks allows us to infer the spatial inhomogeneties in the pattern of interactions across subsystems during this state. The interactions between the two subsystems are most intense in the dump plane but not restricted to this region. The influence of the vorticity dynamics and thermoacoustic power generation in the dump plane is widespread throughout the turbulent reacting flow field of the combustor. Since the hubs in the recirculation zone have dominant disassortative connections to nodes in almost all locations downstream, any perturbation in the hub region region will disrupt the feedback between subsystems and affect the dynamics at downstream locations. Thus, attacking the spatial collection of hubs of the inter-layer network identified in the dump plane is a suggested passive control strategy for mitigating the impending thermoacoustic instability.  
\subsection{Thermoacoustic instability}
\par
\begin{figure}
    \centering
    \includegraphics[width=1\linewidth]{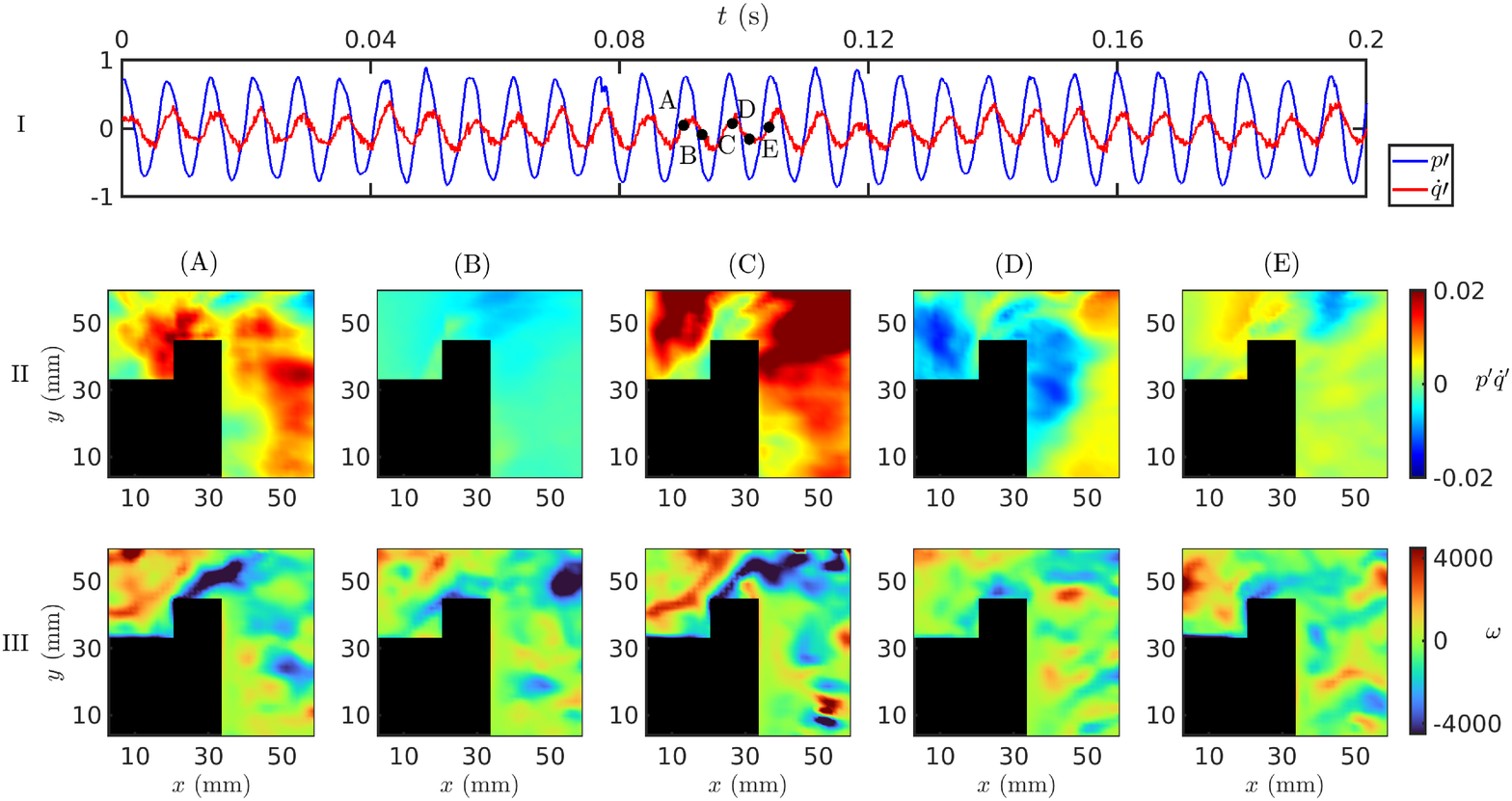}
    \caption{(I)-Time series of acoustic pressure ($p'$) and global heat release rate fluctuations ($\dot{q}'$) during the state of \textbf{thermoacoustic instability}. Spatial distribution of (II)-thermoacoustic power generation and (III)-vorticity at time instants corresponding to points A-D shown in the time series in (I). Coherent thermoacoustic power generation occurs periodically in large regions in the wake and in the dump plane. Large positive vorticity patches appear periodically in the dump plane indicating the periodic shedding of large vortices in this region.}
    \label{fig_raw_765}
\end{figure}

 Order is established in the spatio-temporal dynamics of the combustor during the state of thermoacoustic instability. We obtain high-amplitude periodic oscillations in the acoustic pressure and heat release rate signals (figure \ref{fig_raw_765}-(I)). Coherent heat release occurs and $p'\dot{q}'$ is generated in large regions of the combustion chamber periodically (see figure \ref{fig_raw_765}-(II)A,C). Such sources of high thermoacoustic power occur predominantly in the recirculation zones in the dump plane and the wake of the bluff-body (such as in figure \ref{fig_raw_765}-(II)C). The value of thermoacoustic power generated in the combustion chamber during thermoacoustic instability (refer figure \ref{fig_raw_765}-(II)) is an order of magnitude higher than that obtained during periodic epochs of intermittency (refer figure \ref{fig_raw_570_periodic}-(II)), and about three orders of magnitude higher than that during the state of combustion noise (refer figure \ref{fig_raw_480}-(II)). Also, strong thermoacoustic power sinks are formed in the combustor during such dynamics (see figure \ref{fig_raw_765}-(II)D). 
\par The vorticity field during thermoacoustic instability (figure \ref{fig_raw_765}-(III)) exhibits periodic spatio-temporal patterns, and the magnitude of the vorticity fluctuations are higher than those during the states of combustion noise or during aperiodic epochs of intermittency. We can observe a large region of high positive vorticity in the recirculation zone in the dump plane (see figure \ref{fig_raw_765}-(III)A). From experiments, we observe that very large vortices are formed upstream of the bluff-body during the occurrence of thermoacoustic instability; these vortices span the dump plane and the outer shear layer regions. These large coherent structures are shed periodically at a frequency equal to the acoustic frequency \citep{george2018pattern} and subsequently impinge on the walls of the combustor or the bluff-body. Impingement of these large coherent structures and breaking down into smaller vortices leads to sudden heat release in the combustor.
\begin{figure}
    \centering
    \includegraphics[width=1\linewidth]{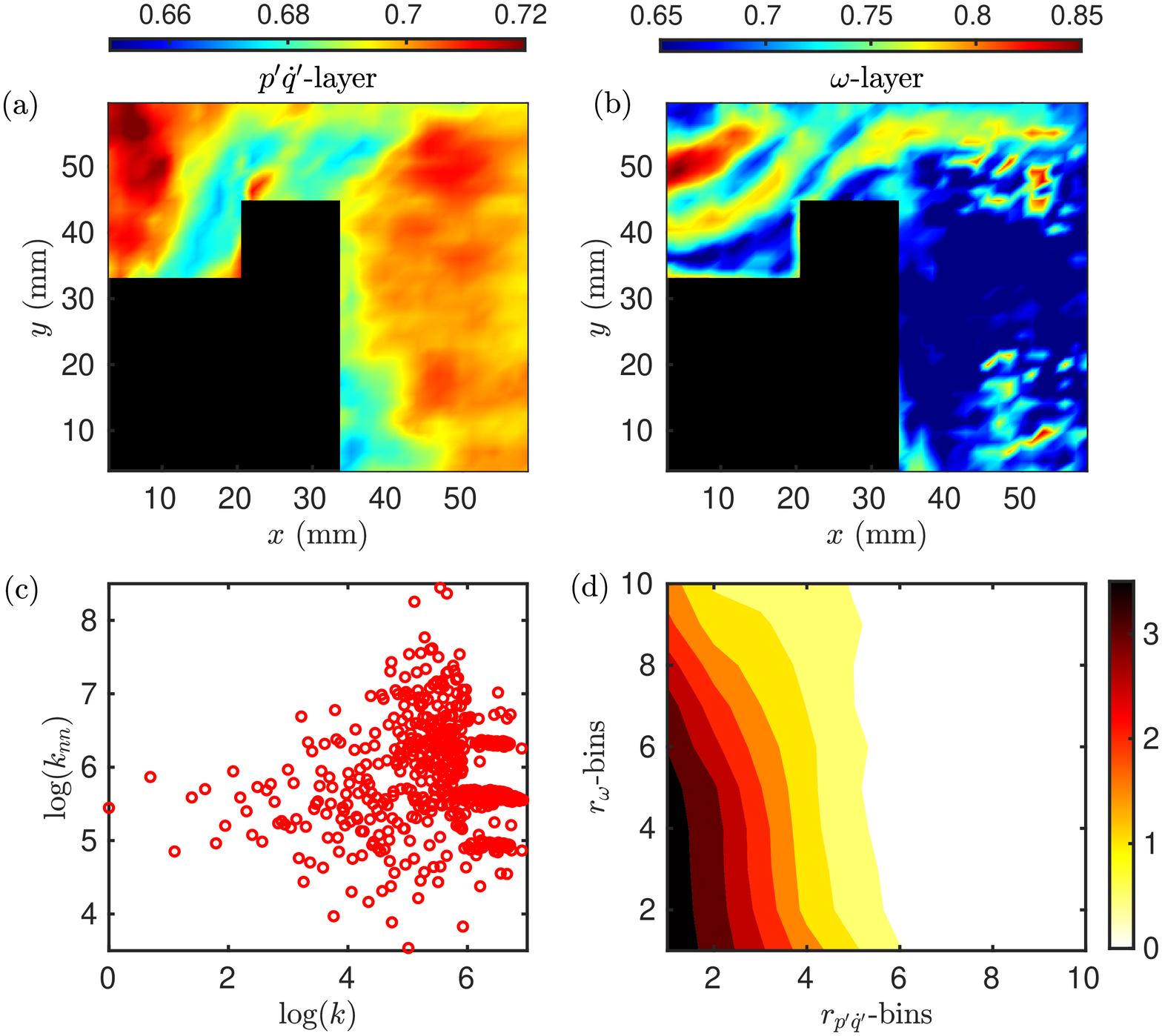}
    \caption{The spatial distribution of node strengths in the (a) $p'\dot{q} '$-layer and (b) $\omega$-layer derived from the inter-layer network during the state of \textbf{thermoacoustic instability}. (c) Variation of the inter-layer degree correlation function ($k_{nn}$) with inter-layer degree ($k$), where the network exhibits neutral assortativity. (d) The inter-layer link-rank distribution of links between nodes ranked into bins in the $p'\dot{q}'$-layer and the $\omega$-layer. The colorbar indicates the percentage of links. Almost all nodes in the dump plane and the wake in the $p'\dot{q}'$-layer are low-ranking nodes (high-node strengths). These nodes predominantly connect to low-ranking nodes in the $\omega$-layer that occur in the shear layer regions and a small patch in the dump plane.}
    \label{fig_765_ILN}
\end{figure}
\par Figure \ref{fig_765_ILN} shows the results from the analysis of the inter-layer network for the state of thermoacoustic instability. In the ${p'\dot{q}'}$-layer (figure \ref{fig_765_ILN} (a)), the node strengths are moderately high throughout the combustion chamber (in the range 0.65 to 0.7). The node strengths are highest in the dump plane followed by high node strengths in a large patch in the wake region of the ${p'\dot{q}'}$-layer. In the $\omega$-layer, we obtain the highest node strengths (around 0.85) in a patch in the recirculation zone in the dump plane, and high node strengths (around 0.75) in the outer and inner shear layer. Nodes in the wake region have moderately high node strengths (around 0.65).
\begin{figure}
    \centering
    \includegraphics[width=0.9\linewidth]{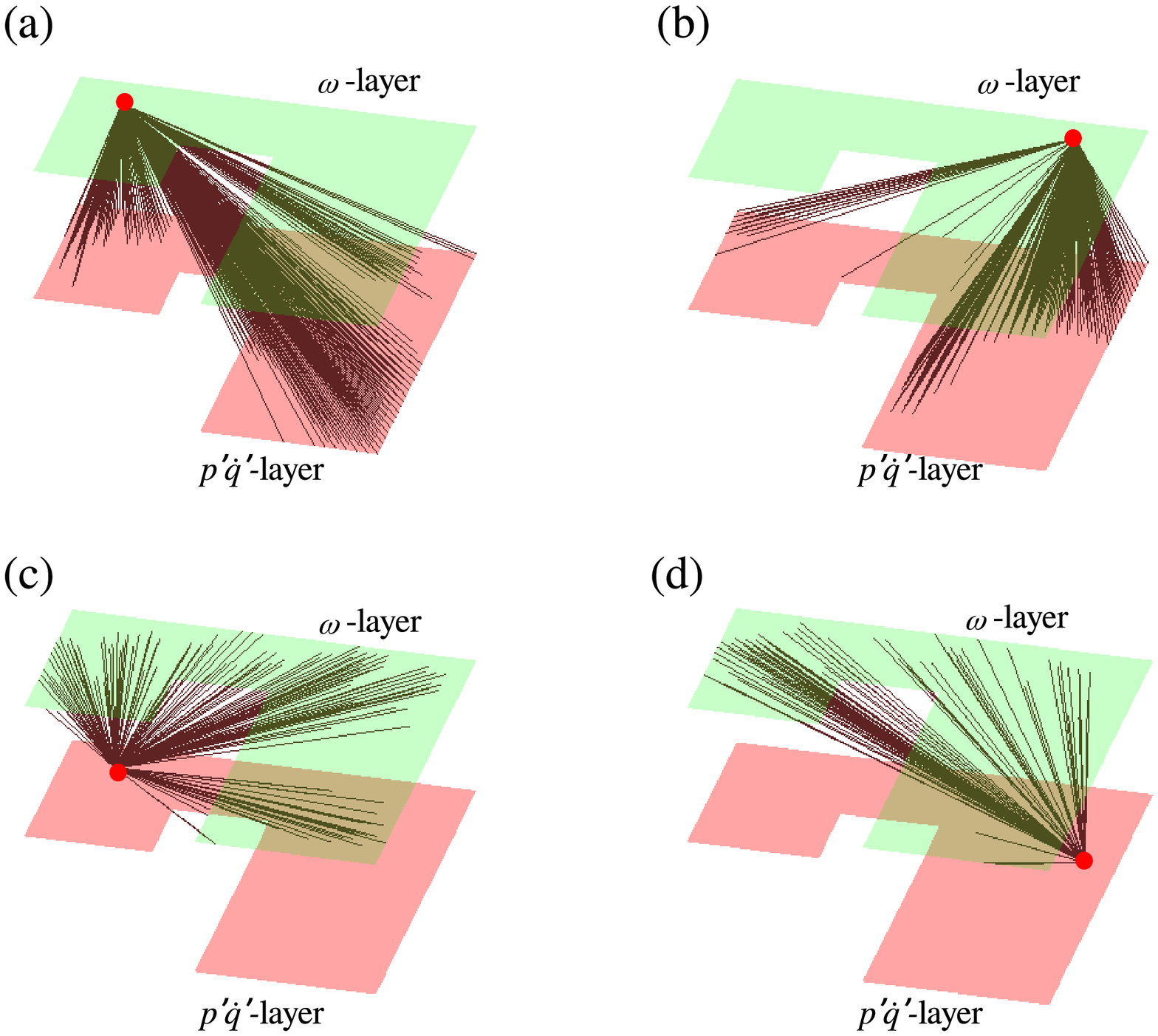}
    \caption{Visualization of inter-layer links derived from inter-layer network during the state of \textbf{thermoacoustic instability} for sample nodes. The green surface represents the $\omega$-layer, while the red surface represents the $p'\dot{q}'$-layer. The cut out region in each layer indicates the location of the bluff-body. Sample nodes are selected (a) in the hub region in recirculation zone of the $\omega$-layer, (b) in the farther downstream end of the inner shear layer of the $\omega$-layer, (c) in the recirculation zone of the $p'\dot{q}'$-layer, and (d) a node in the wake region of the $p'\dot{q}'$-layer. Dense inter-layer links exist between the nodes in the shear layer regions of the $\omega$-layer and the nodes in the dump plane and wake regions of the $p'\dot{q}'$-layer. }
    \label{fig_765_visual_ILN}
\end{figure}
Further, we find that the inter-layer degree correlation function ($k_{nn}$) does not exhibit a monotonic increase or decrease with the inter-layer degree ($k$) and it is not possible to find a power law fit to the degree correlations (figure \ref{fig_765_ILN}(c)). The inter-layer network topology is neutral, that is, neither assortative nor disassortative, owing to the fact that almost all nodes in the ${p'\dot{q}'}$-layer and several nodes in the $\omega$-layer have similar values of node strengths. Using the inter-layer link-rank distribution (figure \ref{fig_765_ILN}(d)), we find that inter-layer links are predominantly low-to-low and low-to-high ($r_{p'\dot{q} '}-\text{to}-r_{\omega}$) type (refer schematic in figure \ref{schem_iln_linkrank}). Note that, almost all nodes in the $p'\dot{q}'$-layer have very high node strengths and are thus binned together as low ranking nodes. 

\par From the spatial distribution of node strengths and the link-rank distribution, we infer that the dense inter-layer links exist in the dump plane and the shear layer regions of the $\omega$-layer and are also spread throughout the flow field in the ${p'\dot{q}'}$-layer. This interpretation is supported by figure \ref{fig_765_visual_ILN}, which shows the inter-layer links of sample nodes upstream and downstream of the bluff-body in both the layers. A node in the hub region in the dump plane of the $\omega$-layer (figure \ref{fig_765_visual_ILN}(a)) has inter-layer connections predominantly to the hub region in the dump plane of the ${p'\dot{q}'}$-layer and also some inter-layer links to the wake region. Figure \ref{fig_765_visual_ILN}(b) shows that a hub-node from the $\omega$-layer in the inner shear layer has links predominantly to the wake region and a few links to the dump plane in the ${p'\dot{q}'}$-layer. Further, a node in the dump plane in the ${p'\dot{q}'}$-layer (figure \ref{fig_765_visual_ILN}(c)) has predominant inter-layer connections to nodes in the dump plane and the shear layer regions of the $\omega$-layer (high node strength regions, see figure \ref{fig_765_ILN}(b)). Also, figure \ref{fig_765_visual_ILN}(d) shows that a node in the wake region in the ${p'\dot{q}'}$-layer has inter-layer connections with nodes in the shear layer regions as well as the recirculation zone in the $\omega$-layer. 

\par From the network topology, we can infer the fluid mechanical processes occurring at different locations in the combustor. The thermoacoustic power generated in the dump plane is strongly correlated to the vorticity dynamics in both the dump plane and the shear layer regions. We conjecture that the large vortices shed in the dump plane enhance both the bulk mixing as well as fine-scale mixing by increasing local strain rates in the fluid \citep{poinsot1987vortex, schadow1992combustion}. Note that, combustion occurs only when reactants are adequately mixed at the fine scales. Since large vortices are formed, adequate fine-scale mixing occurs over larger regions and over longer epochs than during the occurrence of combustion noise or intermittency where relatively smaller vortices were shed erratically in the dump plane. As a result, most of the reactants mix well with the hot products over a certain time period and then undergo combustion at once during the state of thermoacoustic instability. Since a large amount of reactants undergoes combustion simultaneously and suddenly (at the end of the time period of mixing), sudden release of large amounts of heat occurs in the combustion chamber. The heat released due to combustion is periodic owing to the periodic vortex shedding activity in the dump plane. 

The different processes namely, the vortex shedding dynamics, acoustic field fluctuations and combustion leading to sudden heat release, are inter-dependent on each other. These processes co-evolve and synchronize in the dump plane region resulting in periodic spatio-temporal dynamics. Thus, the nodes in the dump plane in the $p'\dot{q}'$-layer have a significant amount of connections to the nodes in the dump plane in the $\omega$-layer. Furthermore, as combustion occurs, the fluid material in the vortices expands and travels downstream. The vortices grow in size due to entrainment of fluid as well as due to combustion and eventually impinge onto the walls of the combustor or on the bluff-body and break into smaller vortices. As a result, the fine-scale mixing is enhanced in the shear layer region which ensures adequate mixing of the leftover unburnt reactants with the hot products. Eventually, these reactants undergo combustion and cause further heat release in the wake region. Thus, the nodes in the wake region in the $p'\dot{q}'$-layer have significant number of connections to the shear layer region in the $\omega$-layer. Therefore, using multilayer network analysis, we are able to describe how the large coherent structures formed during the state of thermoacoustic instability play a pivotal role in fostering inter-subsystem activity. Thus, our findings not only support but also explain the findings of \citet{poinsot1987vortex} that the reaction zone and coherent heat release patches trail behind the large coherent structures during the state of thermoacoustic instability.

We note that, sources and sinks of thermoacoustic power generation occur simultaneously in the wake and dump plane during the state of thermoacoustic instability (see figure \ref{fig_raw_765}-IIC,D). Such simultaneous and periodic occurrence of large spatial patches of positive and negative $p'\dot{q}'$ indicates synchrony between thermoacoustic power generation at spatially separated locations in the flow field. From multilayer network analysis, we understand that the thermoacoustic power generated in the dump plane and the wake are not independent activities. We observe that activity in the $p'\dot{q}'$-field in, both, the dump plane and in the wake are well correlated with the vorticity dynamics in the dump plane and the shear layer regions. We conjecture that a feedback is established between the thermoacoustic power generated in these different regions due to the inter-subsystem interactions with the vorticity field. In other words, the activity of thermoacoustic power generation in the dump plane and wake become synchronized owing to the spatially extended structure of inter-subsystem interactions as described during the state of thermoacoustic instability. Thus, the inter-subsystem interactions promote synchronized intra-subsystem activity across different locations.

\begin{table}
  \begin{center}
\def~{\hphantom{0}}
\begin{tabular}{>{\centering\arraybackslash}m{2.cm}>{\centering\arraybackslash}m{2.cm}>{\centering\arraybackslash}m{2.cm}>{\centering\arraybackslash}m{3cm}>{\centering\arraybackslash}m{3cm}}

\textbf{Multilayer network analysis} & \textbf{High node strength regions in $p'\dot{q}'$-layer} & \textbf{High node strength regions in $\omega$-layer} & \textbf{Inter-layer network topology} & \textbf{Interpretation} \\
\hline
\textbf{Combustion noise} & Shear layer and wake region  & Dump plane and wake & Inter-layer links are spread across almost all locations in the combustor & No localized pockets of intense inter-subsystem interactions\\
\hline

\textbf{Intermittency}   & Recirculation zone in the dump plane & Recirculation zone in the dump plane & Dense hub-to-hub inter-layer links; also significant links between hubs in one layer and low node strength regions in the other layer & Localized pockets of intense inter-subsystem interactions emerge in the dump plane with significant influence throughout the flow field of the combustor\\
\hline

\textbf{Thermoacoustic instability}   & All regions; predominantly the dump plane & Recirculation zone in the dump plane and shear layer regions & Dense connections exist between nodes in the dump plane and shear layer regions of the $\omega$-layer and dump plane and wake region in the $p'\dot{q}'$-layer, respectively & Intense inter-subsystem interactions occur between regions of coherent thermoacoustic power generation and regions of coherent vortex shedding, predominantly in the dump plane \\
\hline

\end{tabular}
\caption{Summary of results from multilayer network analysis for different dynamical states in a turbulent thermoacoustic system}
  \label{summarytable}
  \end{center}
\end{table}
\par The results obtained from multilayer network analysis during the various dynamical states in a turbulent thermoacoustic system are summarized in table \ref{summarytable}. Using this approach, we are able to infer the spatial inhomogeneties in the pattern of inter-subsystem interactions and infer and distinguish the fluid mechanical processes occurring at different locations in the flow field during different dynamical states.
\subsection{Discussion}
\par The multilayer network approach in the current study explains the mechanisms behind the observations from some previous studies. For example, we are able to identify a critical region of inter-subsystem interactions between thermoacoustic power and vorticity field in the dump plane during the state of intermittency and thermoacoustic instability. Such a critical region was also identified in previous studies investigating the hydrodynamic subsystem independently. For example, a similar critical was region identified by \citet{premchand2019lagrangian} corresponding to the dominant acoustic mode using flow decomposition techniques during the occurrence of thermoacoustic instability. Also, a similar critical region was identified by \citet{abin2021jfmtaira} as the hubs of single-layer vorticity networks during the state of thermoacoustic instability. Further, \citet{roy2021critical} found a similar region in the dump plane where intense turbulent velocity fluctuations occur during the occurrence of thermoacoustic instability. 

\par \citet{abin2021jfmtaira} and \citet{roy2021critical} found significant suppression of thermoacoustic instability even for low flow rates of targeted microjet injections in the recirculation zone behind the bluff-body. However, perturbing any other region such as patches identified by the authors at the top of the bluff-body or in the wake, does not readily aid the mitigation of thermoacoustic instability. While the authors \citep{abin2021jfmtaira,roy2021critical} intended to target regions of intense hydrodynamic activity, they inadvertently perturbed the region of significant inter-subsystem interactions which occurs in the recirculation zone in the dump plane as identified using multilayer networks. Thus, we conjecture that targeted attack in regions where dense hub-to-hub inter-layer connections exist can lead to the mitigation of thermoacoustic instability. Such perturbations would disrupt the strongest feedback interactions between the multiple subsystems such as the hydrodynamic field and the acoustically-coupled combustion dynamics. Thus, the use of multilayer networks potentiates physical insight into the occurrence as well as the mitigation of thermoacoustic instability. 
\section{Conclusions \label{sec_conclusion}}

Turbulent thermoacoustic systems are complex systems wherein the coevolution of the hydrodynamic, acoustic and combustion dynamics are inter-dependent due to nonlinear interactions occurring in a spatially extended flow field. In this work, we investigate such inter-subsystem interactions between the vorticity field ($\omega$) and thermoacoustic power ($p'\dot{q}'$) generation during chaotic, intermittent and ordered dynamics in a turbulent bluff-body stabilized dump combustor using complex networks. We construct a multilayer network comprising two layers, the $\omega$-layer and the $p'\dot{q}'$-layer, where the inter-layer connections are derived from the short-window correlation between the two variables $\omega$ and $p'\dot{q}'$ at any two locations in the flow field. The spatial distribution of node strengths derived from inter-layer networks helps us identify regions that foster inter-subsystem interactions (hubs of the inter-layer network) during various dynamical states. Further, we examine the average tendency of a node in one layer to form links with similar or dissimilar nodes in the other layer of the network using the inter-layer degree correlations. Also, by studying the distribution of links between nodes ranked according to their inter-layer degree (link-rank distribution), we identify if significant connections exist between regions of high node strengths across layers in the network. By examining the topology of connections in an inter-layer network, we infer the inter-dependence between the vorticity and $p'\dot{q}'$ fields during each dynamical state. 

Multilayer network analysis helps us infer and distinguish the spatial inhomogeneties in inter-subsystem interactions and the fluid mechanical processes involved during different dynamical states in a turbulent thermoacoustic system. For example, we describe that if small vortices are shed erratically, they convect the reactant mixture quickly to downstream locations therefore generating thermoacoustic power in these regions. On the other hand, when large coherent structures are formed, these structures revolve in the dump plane and accumulate and mix huge amounts of reactants in that region which undergo combustion at once. Thus, we are able to infer the fluid mechanical and combustion processes that are spatially distinct during different dynamical states.

During the state of chaotic spatio-temporal dynamics, we show the inter-subsystem interactions are spread across wide spatial scales due to the small vortices shed erratically. These small vortices introduce wide scales of time delays and cause adequate fine-scale mixing across several locations quickly. These vortices also convect from the dump plane to the shear layer and the wake of the bluff-body easily. The reactants undergo combustion only in the shear layer or the wake region during the state of combustion noise. As a result, the vorticity dynamics in the upstream recirculation zone is found to be strongly correlated with the thermoacoustic power generated at downstream locations in the shear layer and the wake of the bluff-body. 

During the state of intermittency, we discover that inter-subsystem interactions emerge as strong and localized feedback in the dump plane; however, interactions also occur between the dump plane and the downstream locations. Localised inter-subsystem interactions occur due to simultaneous thermoacoustic power generation and shedding of large vortices during the periodic epochs of intermittency. Further, the interactions between upstream and downstream locations occurs owing to the small vortices shed during aperiodic epochs of intermittency or left over reactant mixture from the larger vortices which convects and combusts at downstream locations. 

During the state of thermoacoustic instability, such interactions occur predominantly in the dump plane. Very large coherent structures are formed in the dump plane which enhance both bulk and fine scale mixing. These large vortices delay the fine-scale mixing and increase bulk mixing in the dump plane causing any small vortices to revolve within the dump plane region. As a result, these small scale vortices organize into a large coherent structure \citep{george2018pattern, raghunathan2020multifractal} and do not convect downstream as easily. Thus, after sufficient mixing has occurred, a significant amount of the reactants undergo combustion at once in the dump plane itself. Also, large vortices convect via shear layer region and break down into smaller ones or impinge onto the wall of the combustor causing combustion of the remaining reactant mixture. Hence, we find inter-subsystem interactions are significant between the wake and shear layer regions as well during this state.

In summary, using the approach of multilayer networks, we reveal the rich topology of inter-subsystem interactions in turbulent thermoacoustic systems during various dynamical states. Using this approach, we discover that intense and localised inter-subsystem interactions emerge during the state of intermittency much prior to the onset of order in the spatio-temporal dynamics of the combustor. The acoustic, combustion and vorticity dynamics are inter-dependent during each state; however, such the strength and spatial pattern of inter-subsystem interactions is distinct during each dynamical state. We conjecture, through our analysis, that spatio-temporal order emerges in a turbulent thermoacoustic system when mutually dependent dynamics ensues in localized pockets in the flow field. 
\backsection[Acknowledgements]{We would like to acknowledge Mr. Manikandan Raghunathan, Mr. Nitin George, Dr. Vishnu R. Unni, Mr. Midhun P., and Mrs. K. V. Reeja for providing the experiment data, and Mr. Praveen Kasthuri for the fruitful discussions. We also express our gratitude to Dr. Mahesh Panchagnula for providing the high-speed camera (Photron FASTCAM SA4).}
\backsection[Funding]{This work was funded by the Office of Naval Research Global (ONRG) (grant number N629092212011). T. S. acknowledges the support from Prime Minister Research Fellowship, Government of India.}
\backsection[Declaration of interests]{The authors have no conflict of interest.}
\backsection[Data availability]{The data that support the findings of this study are available from the corresponding author upon reasonable request.}

\appendix
\section{Effect of the size of the window used for computing short-window correlations\label{App_sizewindow}}
To determine links between two nodes in a multilayer network, we have used short-window correlations between the time series of vorticity and the thermoacoustic power at the locations corresponding to the nodes. The size of the window is equivalent to $k_w$ cycles of the dominant mode of acoustic pressure oscillations. Here, we analyse the effect of the size of the short-window used for calculating the $SWA$ correlation. To do so, we examine the structure of the multilayer network during the state of intermittency when $k_w$ is varied from four to ten. 

\begin{figure}
    \centering
    \includegraphics[width=1\linewidth]{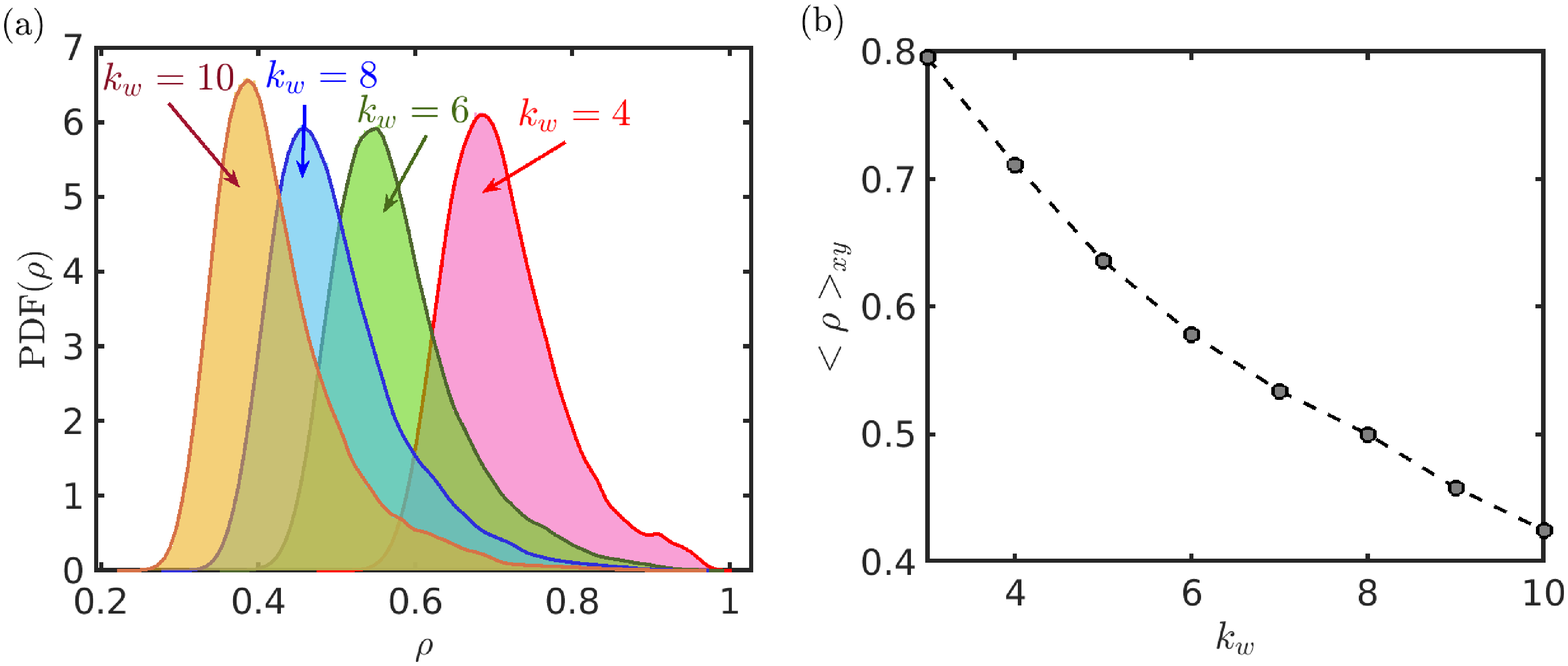}
    \caption{(a) Probability density of short-window adjusted correlations obtained during the state of intermittency for different window sizes equivalent to $k_w$ number of cycles of the dominant acoustic mode. (b) Variation of the spatial mean of correlation ($<\rho>_{xy}$) with $k_w$.}
    \label{windowsize_distcorr}
\end{figure}

Figure \ref{windowsize_distcorr}(a) shows the probability density of $SWA$ correlations between any two spatial locations in the combustor for different values of $k_w$ during the state of intermittency. Clearly, the distribution of correlation values shifts to lower ranges as the value of $k_w$ is increased. Moreover, the spatial mean of the $SWA$ correlations ($<\rho>_{xy}$) decreases rapidly as the size of the short-window is increased (see figure \ref{windowsize_distcorr}(b)) owing to the underlying turbulent flow field. We find a similar trend of decreasing correlations with window size during other dynamical states as well.

\begin{figure}
    \centering
    \includegraphics[width=0.9\linewidth]{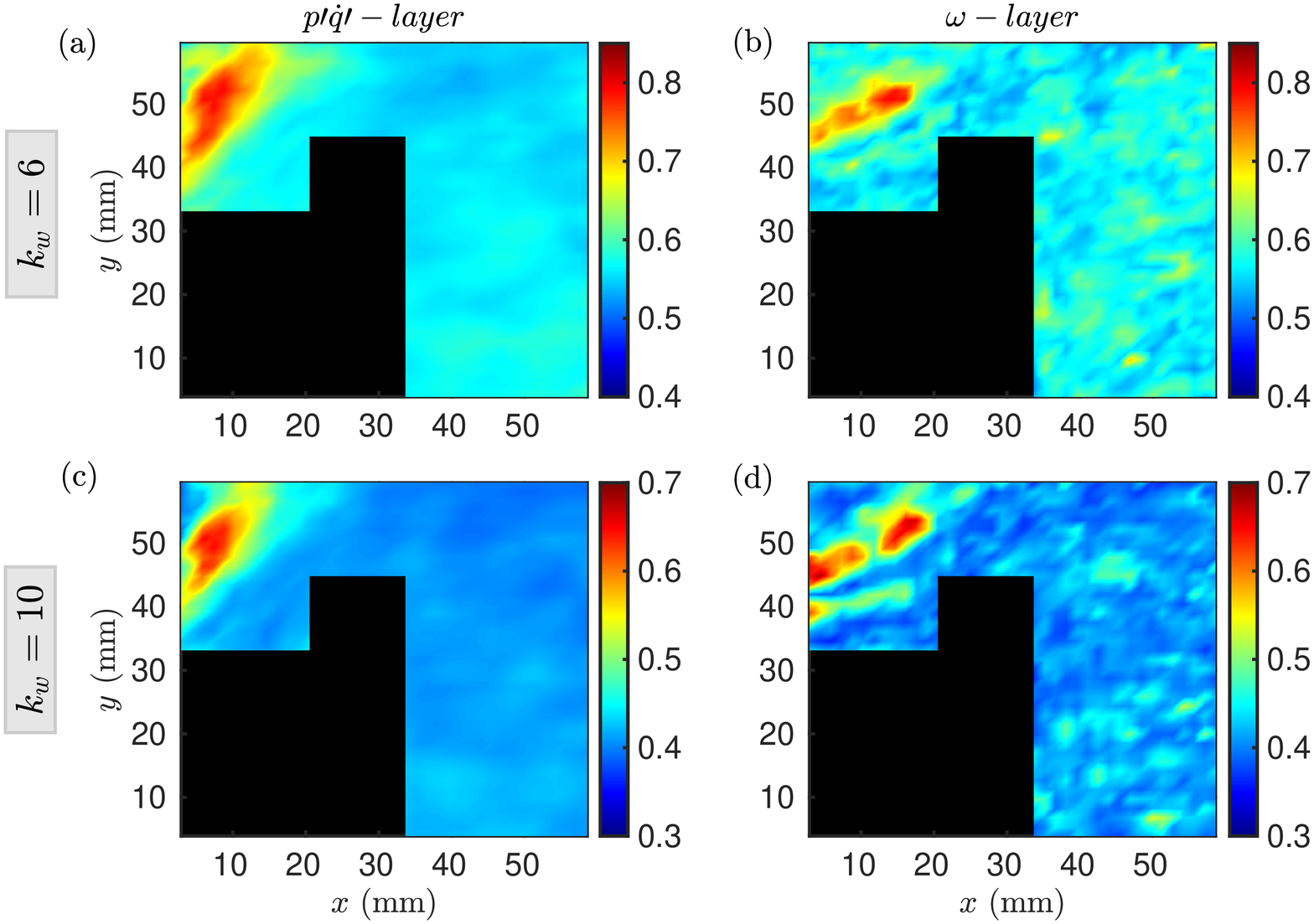}
    \caption{Spatial distribution of node strengths in the $p'\dot{q}'$- and $\omega$ layers obtained during the state of intermittency for $k_w=6$ in (a,b) and $k_w=10$ in (c,d) respectively.}
    \label{windowsize_ns_int}
\end{figure}

Figure \ref{windowsize_ns_int} shows the spatial distribution of node strengths in the $p'\dot{q}'$- and $\omega$- layers of the multilayer network constructed for two different values of $k_w$ during the state of intermittency. We obtain a spatial pocket of hub-nodes in the dump plane in both the layers when the network is constructed using $k_w=6,~10$ (figure \ref{windowsize_ns_int}), which is similar to that obtained in figure 9 where $k_w=4$. Clearly, the spatial distribution of node strengths is robust to the window-size used for computing $SWA$ correlations. In other words, varying the size of the window used for computing $SWA$ correlations affects the range of correlation values but not the spatial distribution of node strengths in the multilayer network obtained from these correlations.

\section{Effect of correlation threshold on degree correlations and link-rank distributions of inter-layer network\label{App_corrthreshold}}

\begin{figure}
    \centering
    \includegraphics[width=1\linewidth]{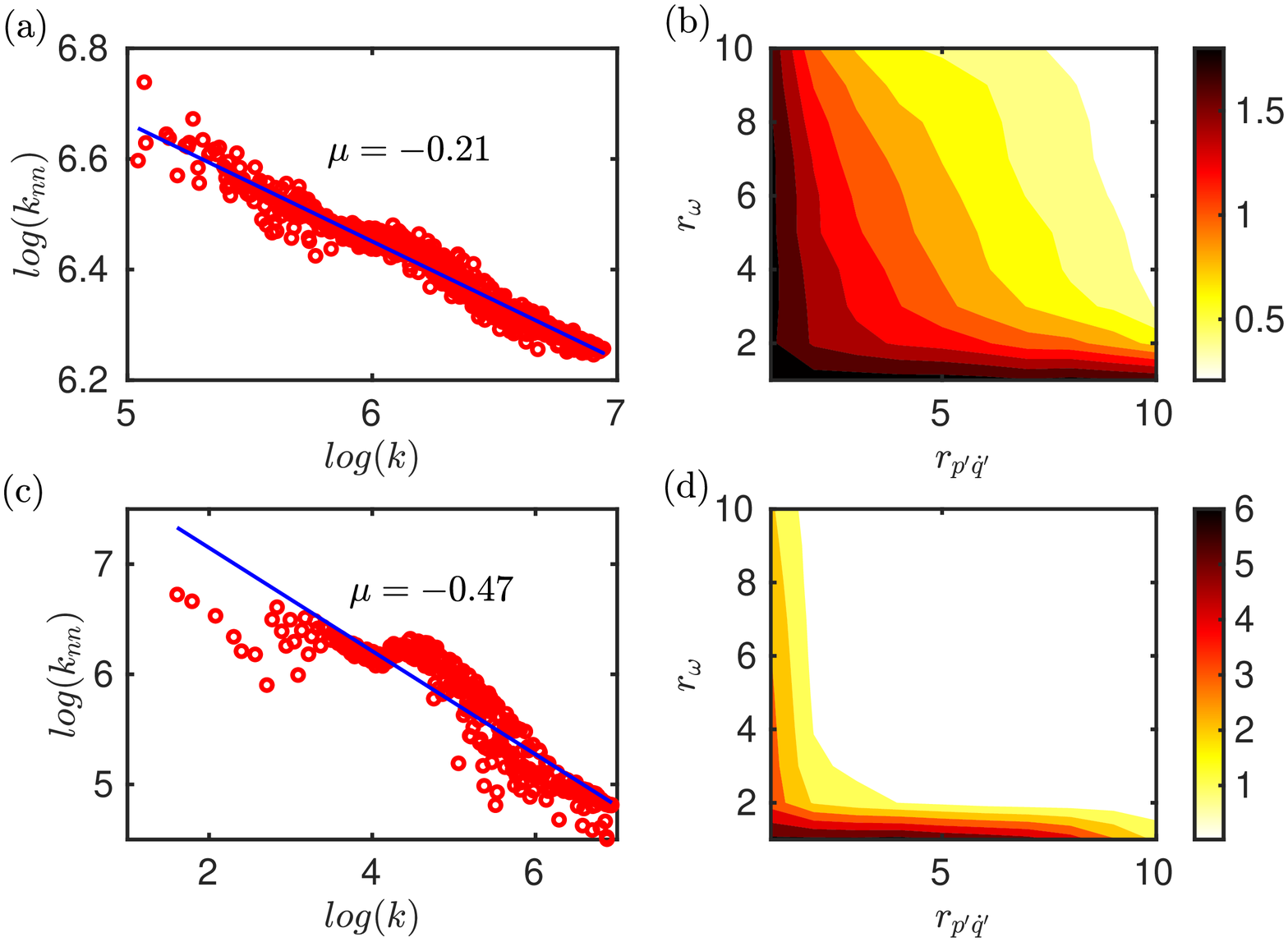}
    \caption{Variation of the inter-layer degree correlation function ($k_{nn}$) with inter-layer degree ($k$) and the link-rank analysis obtained during the state of intermittency when correlation threshold is set at $\rho_{th}= 0.7$ in (a,b) and $\rho_{th}= 0.8$ in (c,d), respectively.}
    \label{fig_degcorr_diff_th}
\end{figure}

In order to examine the distribution of the inter-layer connections with very high weights, we set a threshold on the $SWA$ correlations and obtained an unweighted inter-layer network as explained in Section 4 of the manuscript. Here, we examine the role of this threshold in interpreting the topology of the network. We examine the degree correlations and link-rank distributions obtained from inter-layer network during the state of intermittency for multiple values of the correlation threshold as shown in figure \ref{fig_degcorr_diff_th}. We find that the specific value of correlation exponent changes when the threshold is changed. For $\rho_{th}= 0.75$, used in the manuscript in figure 9(c), the link density of the network was $25\%$. The inter-layer connections exhibited negative degree correlations with a correlation exponent $\mu = -0.39$. 

If we decrease the threshold to $\rho_{th}= 0.75$, the link density of the network increases to $50\%$ and the correlation exponent drops to $\mu = -0.21$ (figure \ref{fig_degcorr_diff_th}(a)). Similarly, when $\rho_{th}= 0.8$, we obtain greater negative slope in the degree correlations ($\mu = -0.47$). Irrespective of the value of the correlation threshold, the network topology remains predominantly disassortative. On increasing $\rho_{th}$, the negative slope of degree correlations increases, implying that the links with higher weights have a greater average tendency of disassortative connections between nodes across layers.  

Further, the link-rank distribution shows that the percentage of links connecting low-ranking nodes across the two layers of the network remains significantly high for any given correlation threshold. Although, the specific distribution of links between node-bins with different ranks does change, the link-rank distribution remains qualitatively same for different correlation thresholds as evident from figure \ref{fig_degcorr_diff_th}(b,d).

\begin{figure}
    \centering
    \includegraphics[width=0.5\linewidth]{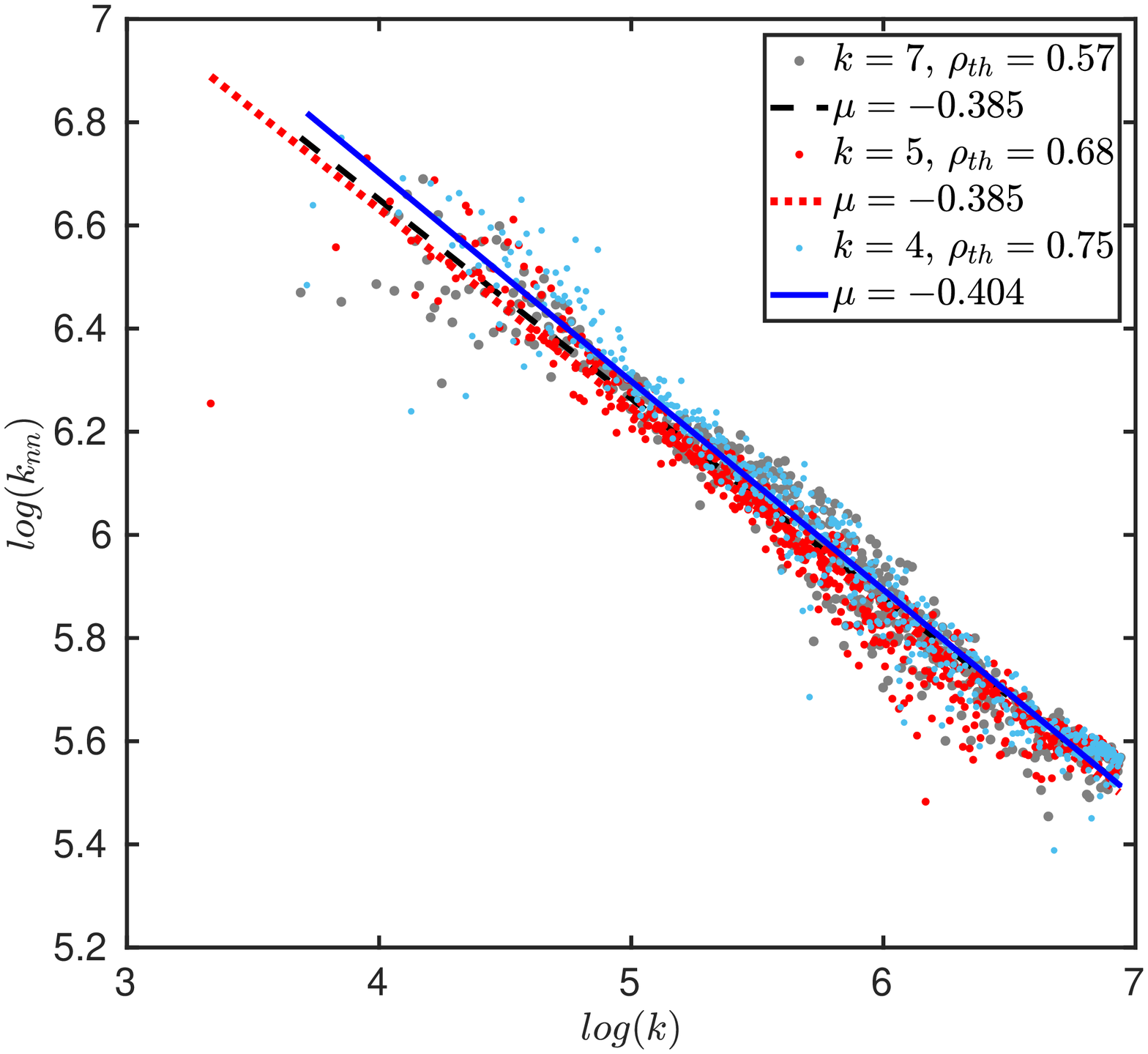}
    \caption{Variation of the inter-layer degree correlation function ($k_{nn}$) with inter-layer degree ($k$) during the state of intermittency when the inter-layer network is obtained using different window-size ($k_w$). For each case of $k_w$, the value of correlation threshold $\rho_{th}$ is chosen such that the link density of the unweighted inter-layer network is $25\%$.}
    \label{fig_degcorr_diff_k}
\end{figure}

In figure \ref{fig_degcorr_diff_k}, we examine how the degree correlations vary when the inter-layer network is constructed using different size of windows ($k_w$) to compute the $SWA$ correlations during the state of intermittency. The correlation threshold $\rho_{th}$ used to extract the unweighted inter-layer network for a given $k_w$ is set such that the link density of the network remains $25\%$. Clearly, the network topology is predominantly disassortative during the state of intermittency irrespective of the size of window used to copute correlations. Moreover, for different $k_w$, but fixed link density, we find that $\mu \approx -0.4$, as evident from figure \ref{fig_degcorr_diff_k}.

\section{Effect of normalising the short-window correlations \label{App_adjcorr}}

Here, we examine the effect of using rematch values to normalise the short-window correlations as explained in Section 3 of the manuscript. The size of the window used to compute correlations is set equivalent to four cycles of the dominant mode of acoustic pressure oscillations ($k_w=4$). We refer to the short-window correlations obtained without normalization as the original values ($\rho_{orig}$), while the correlation values obtained after normalization are referred to as the adjusted values ($\rho_{adj}$). 
\begin{figure}
    \centering
    \includegraphics[width=1\linewidth]{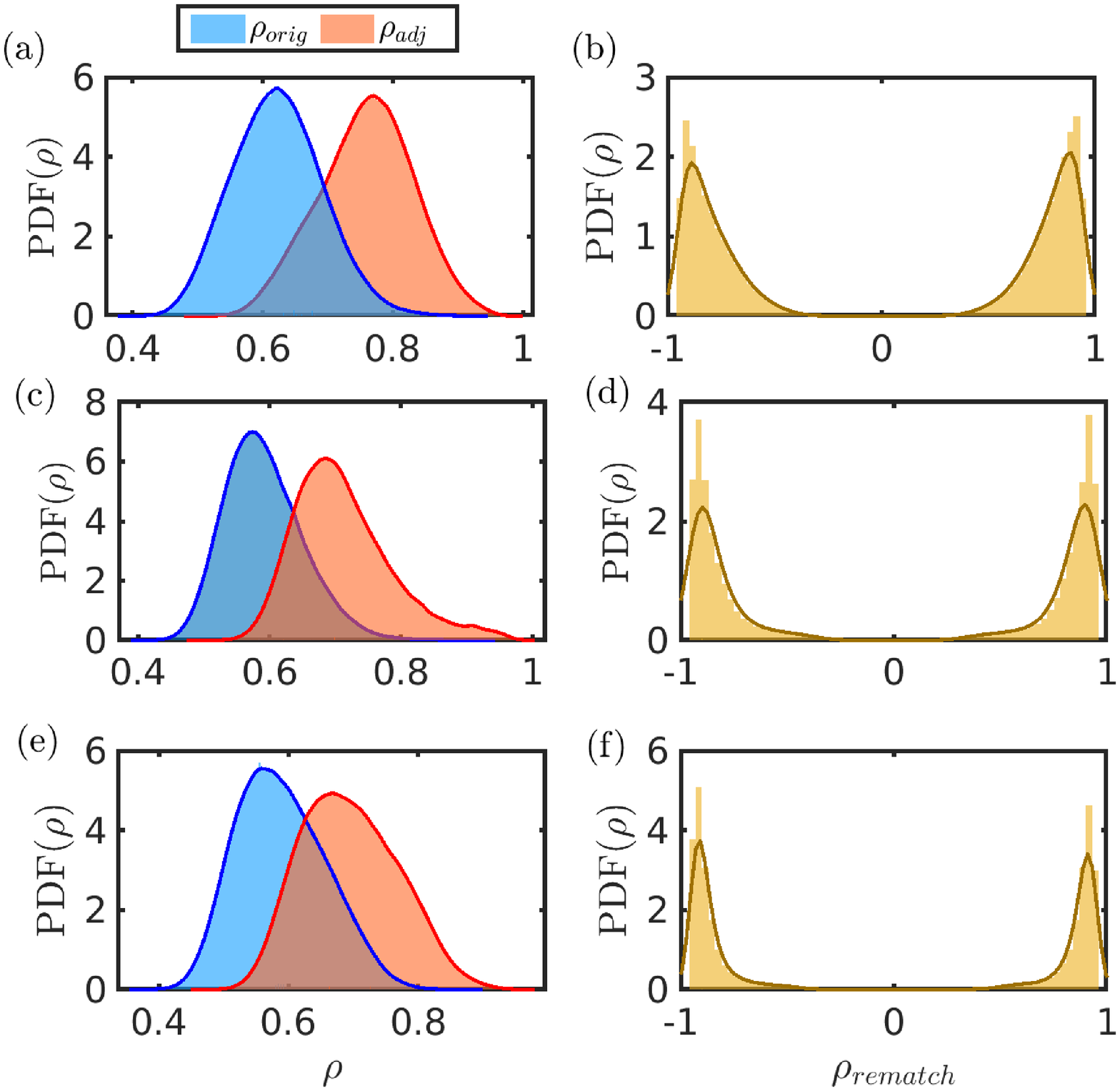}
    \caption{Probability density of original ($\rho_{orig}$) and adjusted ($\rho_{adj}$) short-window correlations during the state of (a) combustion noise, (c) intermittency and (e) thermoacoustic instability. Probability density of the positive and negative rematch values used to normalise the original short-window correlations during the state of (b) combustion noise, (d) intermittency and (f) thermoacoustic instability.}
    \label{orig_adj_prob}
\end{figure}
Figure \ref{orig_adj_prob} shows the probability density of the original (blue curve) and adjusted (orange curve) short-window correlations in the left column, and the probability density of the rematch values used for normalization in the right column during each dynamical state. We find that the original correlation values are significantly lower than the adjusted values during each dynamical state. For example the range of short-window original correlations during the state of thermoacoustic instability is $[0.36,~0.89]$, while due to normalization the range of the short-window adjusted correlations is $[0.46,~ 0.97]$. 

\begin{table}
\centering
\begin{tabular}{ll|ll|ll}
Dynamical state &
\multicolumn{2}{c}{Positive rematch} &
\multicolumn{2}{c}{Negative rematch} &
{Mean $<\Delta \rho>_{xy}$} \\
  & Mean & Mode & Mean & Mode &\\
\hline
Combustion noise  & 0.80 & 0.90 & -0.80 & -0.91  & 22\%\\
\hline
Intermittency & 0.84 & 0.93 & -0.84 & -0.93 & 20\%\\
\hline
Thermoacoustic instability & 0.86 & 0.93 & -0.87 & -0.93  & 18.5\%\\
\end{tabular}
\caption{Statistics of rematch values used to obtain short-window adjusted correlations.}
\label{table_orig_adj}
\end{table}

We have tabulated the mean and mode of the positive and negative rematch values used for normalization during each dynamical state in table \ref{table_orig_adj}. We find that the range of correlations obtained from experimental data is shorter than the theoretically expected range of $[-1,1]$. The maximum and minimum possible cross-variable correlation (positive and negative rematch) is clearly distinct for different pairs of locations. For example, during the state of thermoacoustic instability the rematch values of cross-variable correlations for most of the data are around $\pm 0.93$ (as signified by the mode in table \ref{table_orig_adj}), although for other data points the range of correlations is different leading to a mean of $-0.87$ and $0.86$ for the positive and negative rematch values. As a result, normalization of correlations to achieve a common range is necessary. Also, we report $<\Delta \rho>_{xy}$, the spatial mean of the increase in correlation values achieved by normalization with rematch values. Here, $\Delta \rho = (\rho_{adj}-\rho_{orig})/\rho_{orig}$. Clearly, the use of adjusted correlation values enhances the magnitude of correlation values by approximately $20\%$ during each dynamical state. 

\begin{figure}
    \centering
    \includegraphics[width=1\linewidth]{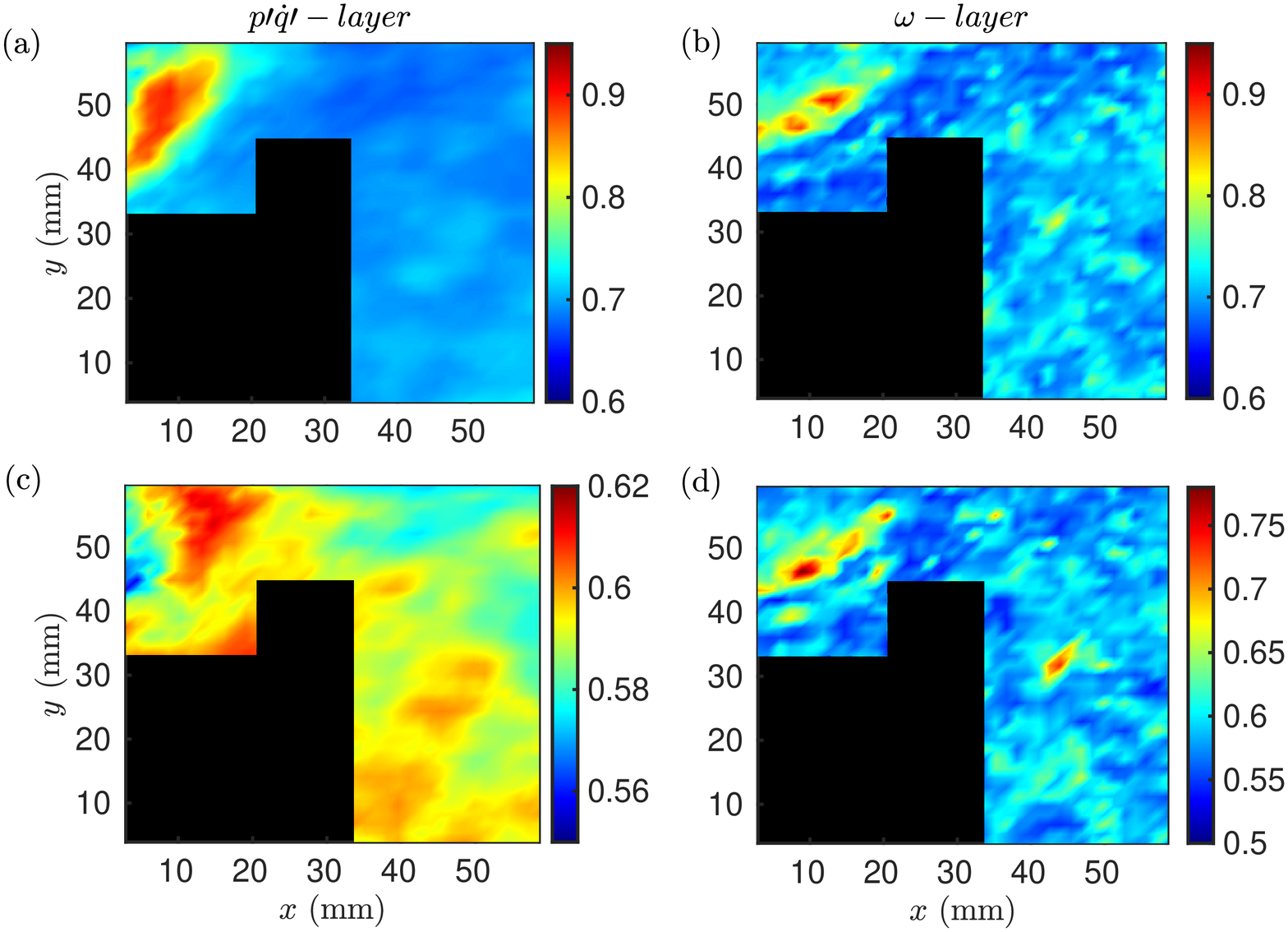}
    \caption{Spatial distribution of node strengths in the $p'\dot{q}'$ and $\omega$ layers for an inter-layer network obtained during the state of \textbf{intermittency} using short-window adjusted correlations in (a,b) and short-window original correlations in (c,d), respectively.}
    \label{orig_adj_corr_INT}
\end{figure}

Next, we plot the spatial distribution of the node strengths during the state of intermittency, when we use adjusted and original short-window correlations to construct the inter-layer network in figure \ref{orig_adj_corr_INT}(a,b) and figure \ref{orig_adj_corr_INT}(c,d), respectively. Using original correlations we obtain a region of high node strength in the recirculation zone in the dump plane in both, the $p'\dot{q}'$ and $\omega$ layers. Also, these regions of high node strengths are similar to that obtained from adjusted short-window correlations in both layers as shown in figure \ref{orig_adj_corr_INT}(a,b). However, variations in the value of node strengths are not significant enough when the network is constructed using the original short-window correlations. As a result, it is difficult to define and distinguish high node strength regions in the flow field. For example the node strengths of nodes in the dump plane are close to that of the nodes in the wake of the bluff-body (see figure \ref{orig_adj_corr_INT}(c)). Note that, the variations in the values of original short-window correlation coefficients is significant as evident from the probability distribution (blue curve) in figure \ref{orig_adj_prob}(c); yet, the node strength values do not have significant variations. Moreover, a comparison between the node strength values at different locations is inapt due to the biased unequal range of correlation values in distinct regions. In summary, the use of adjusted correlations facilitates a fair comparison of correlation values between distinct pair of nodes and also enhances the variations in the node strengths in the network.


\providecommand{\noopsort}[1]{}\providecommand{\singleletter}[1]{#1}%

\end{document}